# A Review of $UTe_2$ at High Magnetic Fields

Sylvia K. Lewin, Corey E. Frank, Sheng Ran, Johnpierre Paglione, and Nicholas P. Butch

**Abstract**

Uranium ditelluride ($UTe_2$) is recognized as a host material to unconventional spin-triplet superconductivity, but it also exhibits a wealth of additional unusual behavior at high magnetic fields. One of the most prominent signatures of the unconventional superconductivity is a large and anisotropic upper critical field that exceeds the paramagnetic limit. This superconductivity survives to 35 T and is bounded by a discontinuous magnetic transition, which itself is also field-direction-dependent. A different, reentrant superconducting phase emerges only on the high-field side of the magnetic transition, in a range of angles between the crystallographic *b* and *c* axes. This review discusses the current state of knowledge of these high-field phases, the high-field behavior of the heavy fermion normal state, and other phases that are stabilized by applied pressure.

## 1. Introduction

Although $UTe_2$ has been studied for decades [1-3], its superconducting state was only discovered a few years ago [4, 5], and many outstanding questions remain. The preponderance of evidence points to an unconventional spin-triplet pairing state, in which the spins of the constituent electrons are parallel, as opposed to antiparallel [4-9]. In addition to the experimental rarity of such configurations, there are additional topological ramifications; namely, the guaranteed presence of non-Abelian Majorana states, if time-reversal symmetry is broken. Indeed, experimental evidence indicates that this may be the case [10, 11], although the identity of the superconducting order parameter remains under debate.

In addition to the novelty of the superconducting order parameter, but not unrelated to it, is the very unusual and dramatic behavior of $UTe_2$ in magnetic field. Indeed, it was the large and anisotropic upper critical field $H_{c2}$, well exceeding the paramagnetic limit in every direction, that was the first indication of spin triplet pairing in $UTe_2$ [4, 12]. Whereas the minimum value of $H_{c2}$ = 6 T along the crystallographic *a* axis, the largest $H_{c2}$ = 35 T along the *b* axis. These magnetic field values are remarkable for a superconductor with a transition temperature $T_c \approx 1.8$ K. However, this is not the whole story; another superconducting phase, which is reentrant or induced by magnetic field, exists at fields between 40 T and 65 T [12]. The temperature dependence of this superconductivity is similar to that of the zero-field superconductivity, and reinforces the very unusual nature of this phenomenon: namely, what fundamental processes stabilize superconductivity with such a high ratio of $H_{c2}$ to $T_c$?

This unusual high-field superconductivity is the focus of the current review. Much experimental progress has been made in several short years, sketching out the magnetic field, temperature, and angular limits of these phases [13]. Given the complexity of this parameter space, it is necessary to remind ourselves that we have so far just scratched the surface. Bordering the superconducting phases are other ordered phases, most prominent of which is a first-order magnetic phase boundary that coincides with both the upper critical field of 35 T and—along a different field direction—the lower bound of 40 T of the reentrant superconducting phase [12, 14, 15]. The intricate relationship between these high-field phases is seen in measurements under applied pressure [16-20]. All of this lives in a background of correlated *f*-electron physics, one of the outstanding theoretical challenges in the study of quantum materials.

This review is organized as follows: in Sec. 2, we summarize the known properties of $UTe_2$ at ambient pressure in zero-to-low magnetic fields, including structure and electronic properties; in Sec. 3, we then review the reported high-field behavior of $UTe_2$ at ambient pressure for various orientations of the magnetic field with respect to the crystal structure, and summarize theoretical proposals for its observed high-field superconducting phases; finally, in Sec. 4 we discuss how both the zero-field and high-field phase diagrams evolve upon the application of hydrostatic pressure.

## 2. Basic properties of $UTe_2$ at ambient pressure and low fields

### 2.1 Crystal structure

Uranium ditelluride crystalizes in a body-centered orthorhombic lattice ($UTe_2$-type *Immm*, No. 71) with typical lattice parameters $a$ = 4.1612, $b$ = 16.1277, $c$ = 13.9614 Å [21]. Unlike the other uranium dichalcogenides, which can crystallize as stable $\alpha$-$UX_2$ (No. 130, tetragonal $Pu_5Rh_3$-type) [22] and $\beta$-$UX_2$ (No. 162, orthorhombic $PbCl_2$-type) [23, 24] or metastable $\gamma$-$UX_2$ (No. 189, hexagonal $Fe_2P$-type) [25] morphologies, $UTe_2$ has no known low-pressure polymorphs. Options for substitution are also limited. So far, the only reported ternary or pseudo-binary $UTe_2$-type compounds besides $UTe_2$ itself are the $(U_{0.5}Ln_{0.5})Te_2$ (Ln = Dy, Ho, Tb, Tm) series [26]. Orthorhombic UTeM compounds, such as USTe and USeTe (No. 162, $PbCl_2$-type) exist [27]; however, none crystallize in the $UTe_2$-archetype.

In the $UTe_2$ structure, a single uranium site (4*i*, point symmetry *mm2*) is eightfold coordinated to two distinct tellurium atoms. Te(1) is located on site 4*j* with *mm2* symmetry and Te(2) is located on 4*h* with *m2m* symmetry. The Te(2) atoms form linear chains parallel to the *b* axis. A Zintl-Klemm evaluation of this structure therefore considers Te(2) hypervalent, which would imply mixed valent uranium with possible charge delocalization [28]. On the other hand, perhaps $UTe_2$ was not fully explored until recently as a candidate for delocalized phenomena such as superconductivity because the U-U interatomic distances exceed the Hill limit (3.5 Å) in every direction [29, 30]. As seen in Figure 1, $UTe_8$ bicapped trigonal prism building blocks stack Te(2) square-face to Te(2) square-face along the *c* axis, forming fourfold capped biprisms that face share across Te(1) triangular faces along the *a* axis, the magnetic easy axis (typical U-U distance 4.16 Å) and corner and edge share along the *b* axis, the magnetic hard axis (typical U-U distance 6.13 Å). Though the distance from uranium dimer to uranium dimer along the *c* axis is rather long at 10.20 Å, U-U distance within the dimers is relatively short (3.763 Å) [21, 31, 32]. Considering the large potential spatial extent of *f* orbitals, this distance may not preclude the possibility of U-U 5*f*-5*f* interactions. Indeed photoemission spectroscopy experiments, discussed below, indicate a heavy 5*f* band intradimer bound state with low orbital angular momentum [33].

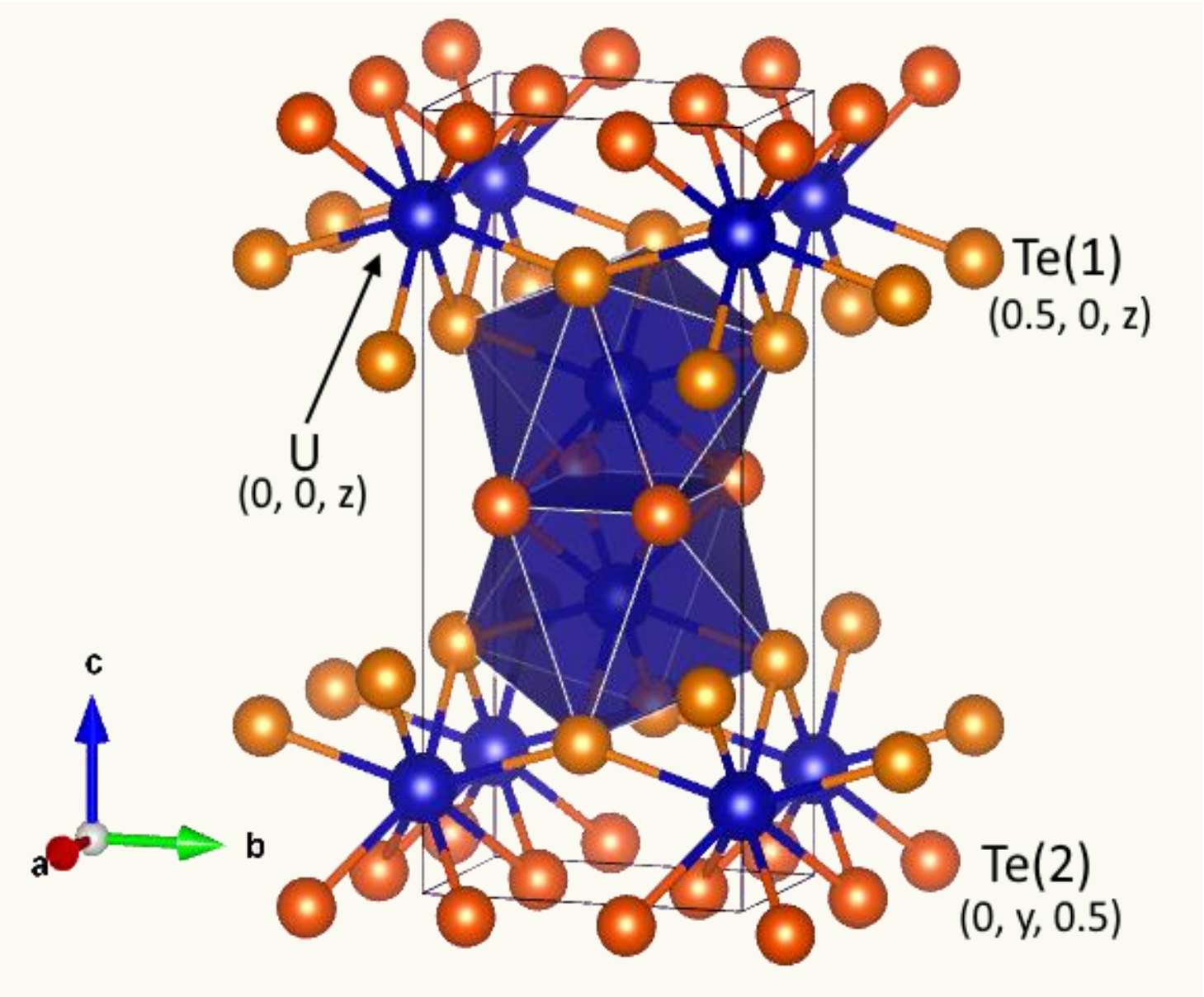

Figure 1. Graphical representation of the crystal structure of $UTe_2$. Uranium atoms are highlighted in blue, Te(1) in gold, and Te(2) in orange. Uranium atoms are eightfold coordinated to four each Te(1) and Te(2), which form the vertices of biaugmented triangular prisms. The coordination polyhedra have been deleted from all but the central pair in this projection to better visualize the fourfold capped biprisms formed by face-sharing dimers in the (001) direction. Figure and typical atomic distances based on data from [32]. The crystal structure was drawn using VESTA software [34].

## 2.2 Low-field magnetization

At low fields, the normal state magnetization of $UTe_2$ exhibits a distinct anisotropy when applied field, H, is parallel to the three crystallographic axes. As shown in Figure 2(b), measurements of magnetization vs field strength (M(H)) on oriented single crystals confirm the *a* axis as the magnetic easy axis at low fields, as perhaps implied by the crystal structure [4]. In magnetization vs temperature (M(T)) experiments, paramagnetic behavior is observed for all three crystal orientations at high temperatures, with Ran *et al.* reporting a somewhat low effective magnetic moment of 2.8 $\mu_B$/U. Previously, Ikeda *et al.* reported an effective magnetic moment of 3.2-3.7 $\mu_B$/U, much closer to the expected magnetic moment for fully localized $5f^2$ or $5f^3$ uranium (3.6 $\mu_B$/U) [3]. These earlier measurements by Ikeda *et al*. did not go below 2 K; it is likely that the measured crystals measured were non-superconducting, as they were grown using a U:Te ratio which has since been observed to consistently produce non-superconducting $UTe_2$.

Both superconducting [4] and likely non-superconducting [3] samples show a broad maximum in M(T) with field along the *b* axis at roughly 30 K. Below this maximum, the magnetization first decreases with decreasing temperature and then becomes temperature-independent, as expected for a coherent Kondo lattice [35]. A much subtler slope change is also observed when H || *a*, at approximately 10 K. Neither of these exhibits the sharp peak in magnetization that would be characteristic of a magnetic phase transition. There is also no accompanying anomaly in specific heat that would indicate a first or second order phase transition, and so these anomalies in M(T) are reasonably attributed to the onset of Kondo coherence.

Below $T_c$, the M(H) of $UTe_2$ has a diamond shape characteristic of the superconducting state; there is also no sign of ferromagnetic ordering in the magnetization even down to 80 mK [36]. Neutron diffraction experiments show no sign of magnetic order down to 2.7 K [31].

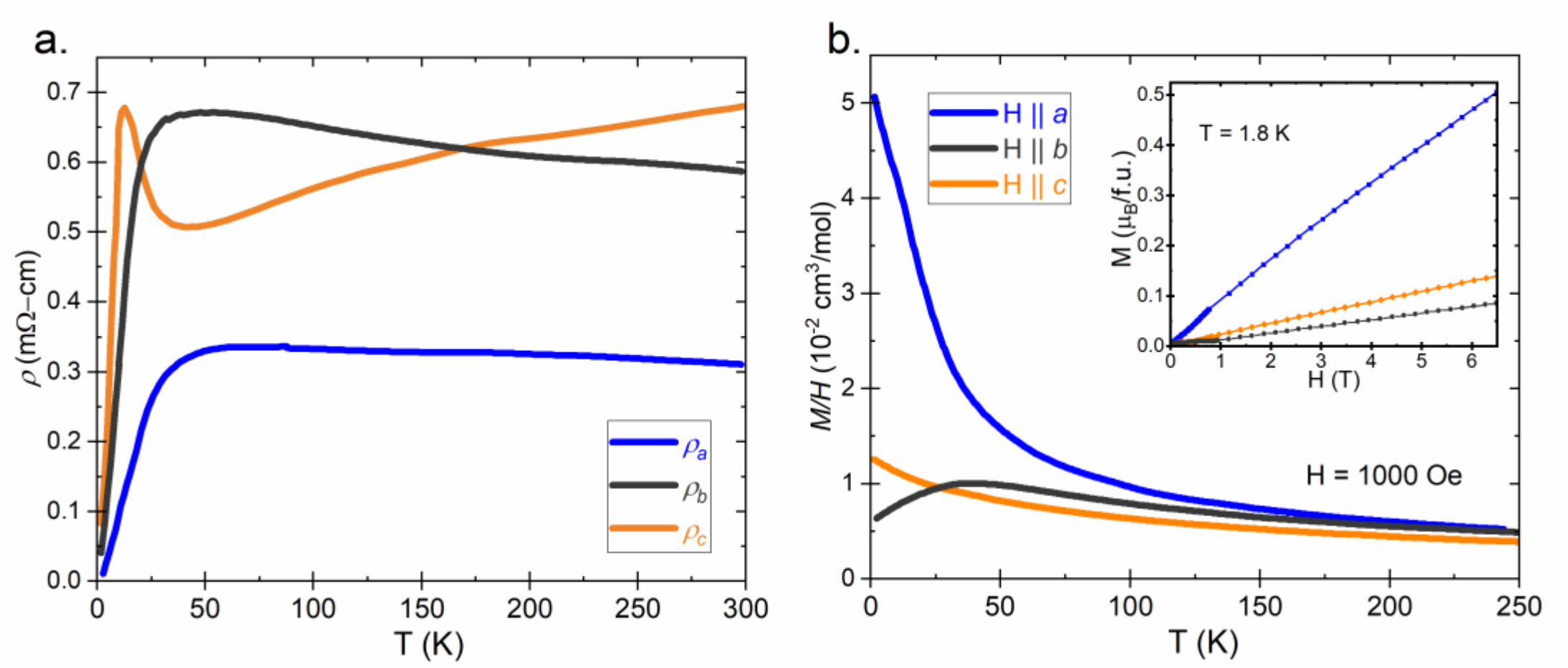

Figure 2. (a) Resistivity of a single crystal of $UTe_2$ in zero field with current applied along the *a* (blue), *b* (black), and *c* (orange) axes, respectively; data from [37]. (b) Magnetization per field vs temperature for a single crystal of $UTe_2$ in the normal state with H applied along the *a* (blue squares), *b* (black diamonds), or *c* (orange circles) axes, respectively. Inset are magnetization vs field data for the corresponding crystal orientations measured at 1.8 K, just above the superconducting transition temperature for this sample; data for figure and inset from [4].

## 2.3 Normal state electrical resistance

Normal state anisotropy is also apparent in the electrical resistivity (ρ) of $UTe_2$, as shown in Figure 2(a). When the current is applied along the *a* or *b* axes, the resistivity has a subtle upward slope as temperature cools from 300 K to about 50 K, which is typical of single-ion Kondo behavior in the decoherent state [37]. Resistivity then decreases rapidly below approximately 50 K, attributed to the onset of Kondo coherence. In contrast, *c*-axis resistance in the high temperature regime *decreases* markedly upon cooling and then a peak is seen at 14 K, below which resistance decreases sharply as it does for the other two principal current directions. This peak has yet to be definitively attributed to a single phenomenon, but possible explanations include a crossover between different conducting channels or decreasing carrier density [37]. Below approximately 5 K, the normal-state resistivity exhibits $T^2$ dependence irrespective of current direction, characteristic of a Fermi liquid [37].

## 2.4 Fermiology

Angle resolved photon emission spectroscopy (ARPES) experiments, coupled with DFT+DMFT calculations, have aimed to elucidate the non-trivial electronic structure of $UTe_2$ by probing its Fermi surface. However, early experiments using soft synchrotron X-rays were complicated by incoherent scattering at the Fermi level, resulting in an anomalous admixture of U 5*f* states (which should be itinerant) into the Te 5*p* bands at deeper binding energies and a non-dispersive band at $E_B$ = 0.5-0.6 eV which calculations at the time could not explain [38].

More recent high resolution ARPES experiments were able to focus on the 5$^{th}$ and 7$^{th}$ Brillouin zones of $UTe_2$, where strong incoherent backgrounds dominate less egregiously [33]. These investigations also revealed a nondispersive band feature at $E_B \approx 0.7$ eV, which Miao *et al.* attributed to the excitations of an atomic multiplet Kondo lattice [33].

The latest ARPES results indicate that the Fermi surface is dominated by quasi-two-dimensional Fermi pockets along $k_z$, which are formed by the hybridization of light quasi-one-dimensional bands [33]. In addition to these quasi-cylindrical pockets, a possibly heavy electron pocket surrounding the Z-point was observed, the dimensionality of which is not yet determined; this pocket could play a role in spin-triplet pairing [33].

In a recent publication, Aoki *et al*. reported the observation of de Haas-van Alphen oscillations in fields up to 15 T [39]. The angle-dependence of the measured quantum oscillation frequencies indicates cylindrical Fermi surfaces along $k_z$; through comparison with calculations, the authors assign the measured frequencies to one electron-like and one hole-like cylindrical Fermi surface.

### 2.5 Zero-field superconductivity

The zero-field superconducting state of $UTe_2$, which we here denote SC1, emerges upon cooling from a paramagnetic state. This is in contrast to other known uranium-based superconductors, such as UCoGe and URhGe, which order ferromagnetically [40]. The SC1 state was first observed in single crystals of $UTe_2$ grown by chemical vapor transport as described in Ref. [4]. These seminal superconducting crystals have a transition temperature of approximately 1.6 K as observed in resistivity, AC susceptibility, and specific heat measurements [4, 5, 12].

The upper critical fields of the SC1 state is highly anisotropic: $H_{c2}(T\rightarrow 0\text{ K})$ is approximately 7 T for H ‖ *a* axis and approximately 11 T for H ‖ *c* axis [5]. Superconductivity persists up to roughly 35 T with field along the *b* axis [16]. However, as shown in Figure 3, this is not necessarily the upper critical field of SC1: there is a superconducting region, SC2, that only appears with a magnetic field applied near the *b* axis. It is not yet unambiguously confirmed whether SC2 is a distinct superconducting phase or simply a field-reinforced region of the SC1 phase, but emerging thermodynamic evidence indicates that SC2 is indeed a separate bulk superconducting phase [41]. Several possible theoretical explanations of the SC2 phase are discussed in Sec. 3.5.

Even the lowest $H_{c2}(0\text{ K})$ of the SC1 phase, approximately 7 T for H ‖ *a*, is well above the paramagnetic (Pauli) limit that constrains conventional spin-singlet superconductors, which should be roughly 3 T based on a 1.6 K transition temperature [4]. This indicates that the superconducting ordering is spin-triplet rather than spin-singlet in nature.

Further evidence for spin-triplet superconductivity comes from nuclear magnetic resonance (NMR) measurements: for a spin-singlet superconductor there should be a significant decrease in the Knight shift below the critical temperature, whereas for a spin-triplet superconductor the Knight shift may either decrease slightly or remain unchanged, depending on the direction of the applied magnetic field relative to the spin components of the spin-triplet pairing. Initial NMR measurements on powder indicated that the $^{125}$Te Knight shift is constant through the critical temperature of the SC1 phase [4], while subsequent measurements showed a slight decrease in the Knight shift when field is along the crystallographic *b* axis or *c* axis and no measurable change along the *a* axis, consistent with spin-triplet pairing [6, 7, 42].

The temperature dependence of the penetration depth and thermal conductivity of $UTe_2$, along with the magnetic field dependence of the thermal conductivity, are consistent with a point-node gap structure [8]. Given the $D_{2h}$ point group of $UTe_2$, this would necessitate a spin-triplet superconducting state. Additionally, scanning tunneling microscopy measurements indicate the existence of chiral states within the superconducting gap [11]. This suggests a topologically non-trivial superconducting state, which could arise from a spin-triplet state with broken time-reversal symmetry.

Broken time-reversal symmetry itself has been measured through the polar Kerr effect in $UTe_2$ [10]. Based on the point group of $UTe_2$ and its strong spin-orbit coupling, the broken time-reversal symmetry below the transition temperature of SC1 indicates a two-component order parameter with the components belonging to two different irreducible representations. Such a multicomponent order parameter is also supported by some specific heat measurements, which reveal two distinct phase transitions as a function of temperature [10].

A double phase transition in heat capacity is observed consistently for crystals grown following the synthesis method of Ref. [4] when measurements are taken at fine enough temperature steps [10]. However, measurements of $UTe_2$ crystals with slightly different growth conditions have only shown a single transition in heat capacity [5, 43, 44]. The transition temperature has also been shown to vary depending on the details of crystal growth, including furnace temperatures and starting atomic ratios of Te to U in the growth [43-45]. Crystals with transition temperatures up to 2 K have been measured, and these only show a single transition in heat capacity [44]. Furthermore, Thomas *et al.* have shown examples of crystals with two transitions due to sample inhomogeneity [46]. Whether a multicomponent order parameter is an inherent property of $UTe_2$ or a product of disorder is a question currently under investigation.

The pairing interactions underlying superconductivity are also a topic of active debate and research. Based on the perceived similarity to the ferromagnetic uranium superconductors, the dominant magnetic susceptibility in $UTe_2$ has been assumed to be ferromagnetic. This theme is consistent with measurements of the optical Kerr effect, which indicate a very magnetically polarizable normal state [47] and muon spin relaxation, which suggest proximity to a ferromagnetic critical point [48] and the presence of short-range magnetic order [49]. However, neither short- nor long-range magnetic order has been identified in neutron diffraction [31] or subsequent neutron experiments. Inelastic neutron scattering measurements suggest that the dominant magnetic fluctuations occur at antiferromagnetic wavevectors [50, 51] along with the evolution of an additional magnetic anomaly in the superconducting state [52, 53]. Whereas this appeared at first to be problematic for spin-triplet superconductivity, theoretical analysis suggests that antiferromagnetic fluctuations can be consistent with a triplet pairing state [54, 55]. However, these fluctuations may not actually indicate antiferromagnetic correlations. It was demonstrated that the energy scale (4 meV) and temperature dependence of these apparently antiferromagnetic fluctuations suggests an origin in the *f*-electron hybridization in $UTe_2$, rather than incipient magnetic order [56]. The magnetic interactions relevant to superconducting pairing likely have lower energy scales than the observed correlations. Nuclear magnetic resonance measurements have been interpreted in the context of a mixture of ferromagnetic and antiferromagnetic interactions [57].

## 3. High-field behavior at ambient pressure

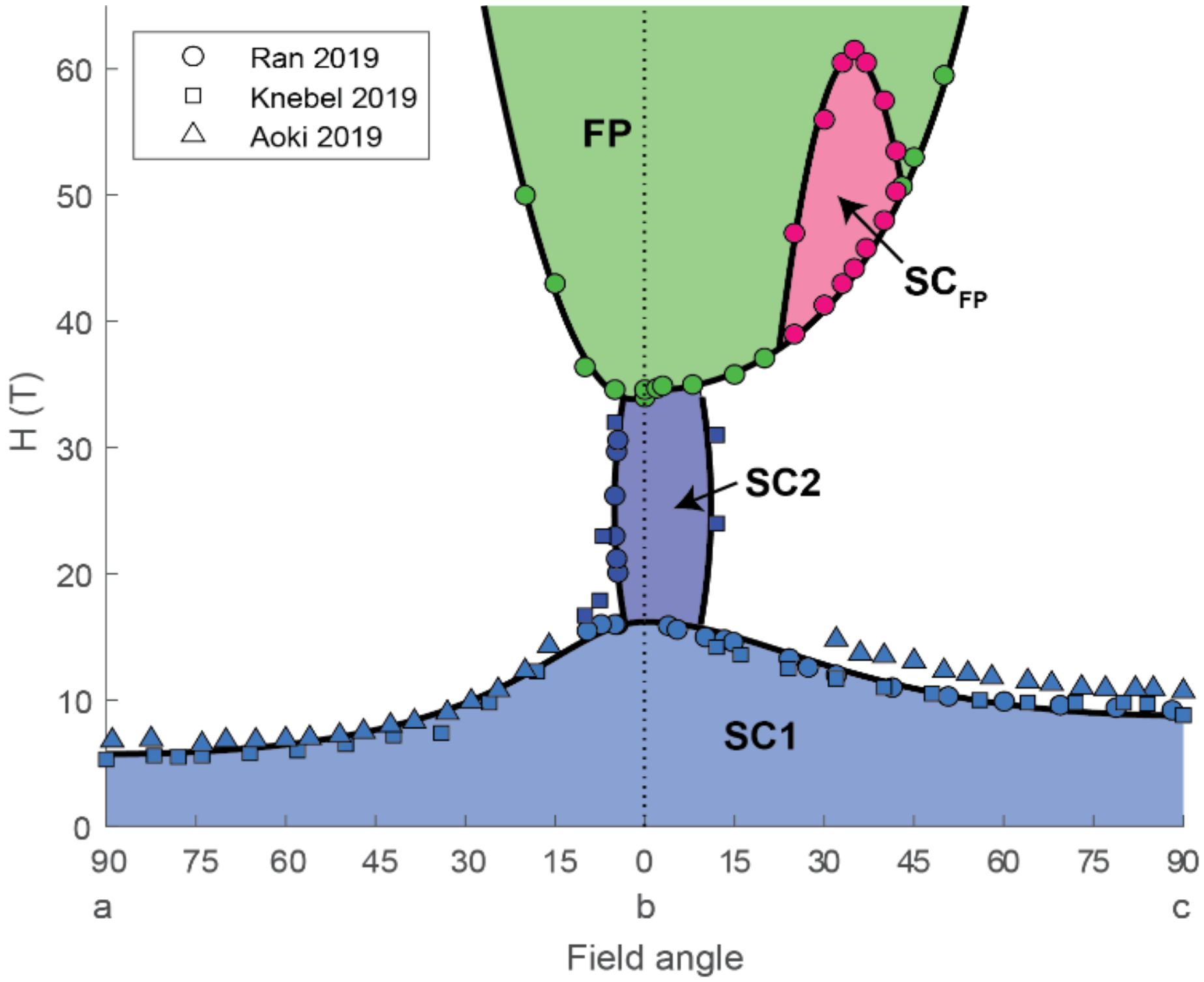

Figure 3. Phase diagram of $UTe_2$ at near-zero temperature, as a function of the magnitude of the applied magnetic field and its angle with respect to the $UTe_2$ crystal axes. The FP phase and the superconducting phases SC1, SC2, and $SC_{FP}$ that are marked on this diagram are all described in the text. Data are from Refs. *[12]* (circles), *[58]* (squares), and *[5]* (triangles).

### 3.1 Superconductivity up to 35 T near *b* axis

The SC2 phase of $UTe_2$ persists up to roughly 35 T and only exists when a magnetic field is applied in a narrow angular range near the crystalline *b* axis, as shown in Figure *3*. The limits of superconductivity are based on measurements of electrical transport [4, 5, 58-60], as shown in Figure 4, and the Seebeck coefficient [59]. Figure 5 shows the evolution of the SC1 and SC2 phases as a function of both temperature and *b*-axis magnetic field. Since resistivity and the Seebeck coefficient are zero in superconductors, these measurements cannot distinguish whether there is a phase transition between SC1 and SC2—and if so, precisely where it lies. However, a recent publication from Rosuel *et al.* shows evidence of a phase transition between SC1 and SC2 in specific heat, supporting the conclusion that SC2 is a distinct phase rather than merely a field-reinforced region of SC1 [41]. This is also evidence of the bulk nature of the SC2 phase, whereas measurements of electrical transport and the Seebeck coefficient cannot easily distinguish a bulk phase from surface or filamentary superconductivity. In additional manuscripts, subtle features have also been reported in proximity diode oscillator, magnetocaloric, AC susceptibility, and NMR measurements that support the presence of a phase transition between the SC1 and SC2 phases [61-63].

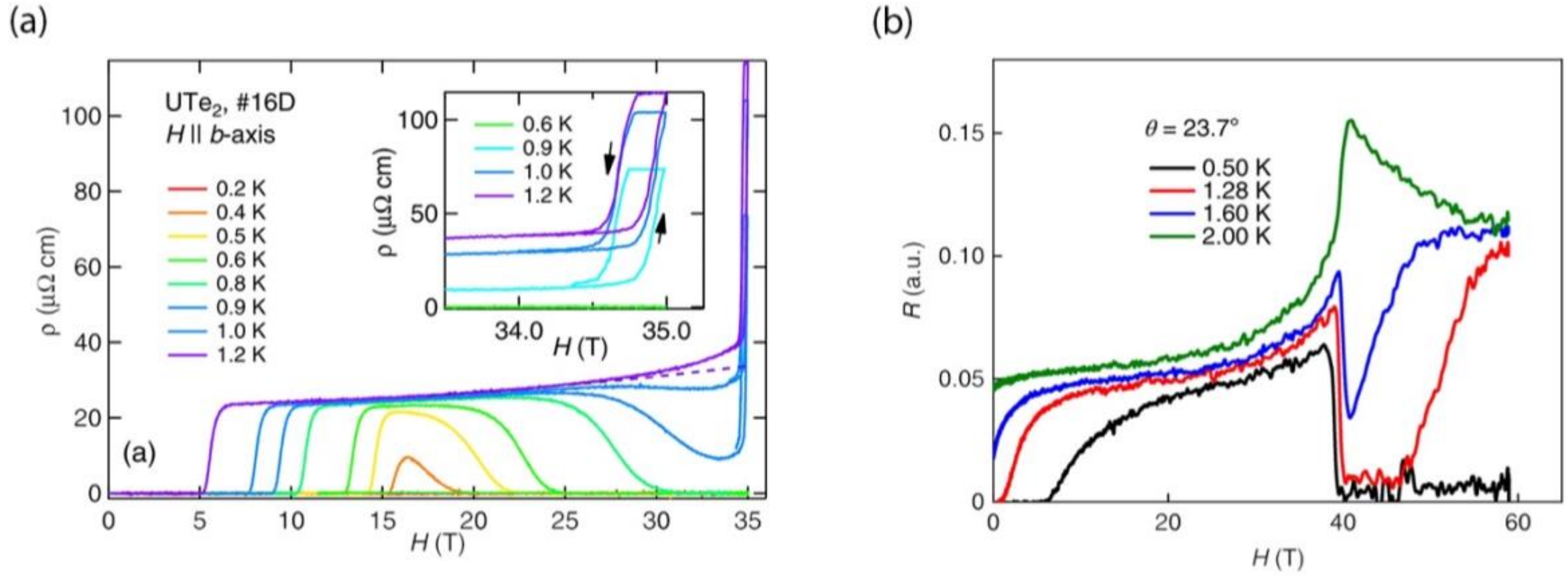

Figure 4. (a) The temperature dependence of the resistivity with field along the *b* axis, from Ref. [58]. (b) The temperature dependence of the resistivity with field tilted approximately 24 degrees from the *b* axis towards the *c* axis, from Ref. [12].

As shown in Figure 5, all of the measurements that extend up to 30 T or higher show an increase of $T_c$ with increasing field (as does Ref. [41]). Thus, SC2 is often referred to as a field-reinforced phase. There is also some discrepancy in the measurements of the phase boundaries of SC1 and SC2, as evident in Figure 5. Given the highly anisotropic behavior of $UTe_2$ in field, it is possible that these differences are caused by unintentional misalignment. They may also be due to variations in sample preparation; note that even in zero field, the superconducting transition temperatures of the samples in Figure 5 are slightly different. Future studies with precise alignments should be able to resolve some of these discrepancies, as well as clearly identify the boundaries of the SC2 phase as a function of angle with respect to the *b* axis.

The gap function of a spin-triplet superconductor can be parameterized by a vector, **d**; for a unitary state it can be proved that **d** is perpendicular to the spin of the Cooper pairs, while for a non-unitary state the physical interpretation of the **d** vector is more complicated [64, 65]. NMR measurements at low and intermediate fields have placed some constraints on the **d** vectors of the SC1 and SC2 phases. Specifically, in the SC1 phase it appears that the **d** vector is strongly pinned along the *b* axis and also has a non-zero component along the *c* axis but no measurable component along the *a* axis [7, 42], while in the SC2 phase the **d** vector has no component along *b* [42, 63, 66]. In terms of the $D_{2h}$ point group symmetry of $UTe_2$, these results indicate that the order parameter of the SC1 phase has a $B_{3u}$ component, while the order parameter in the SC2 phase belongs to $B_{2u}$ [7, 42].

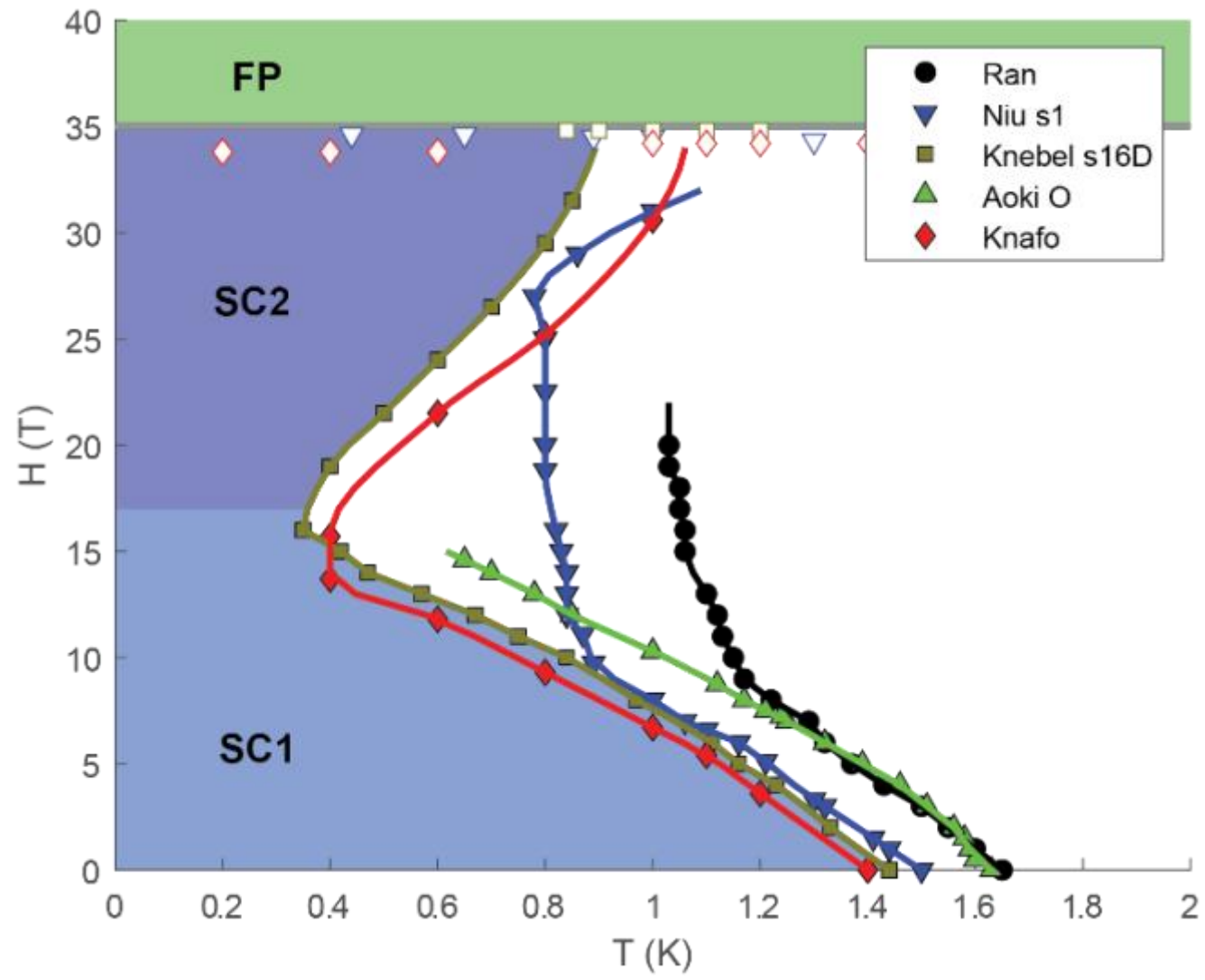

Figure 5. $H_{c2}$ as a function of temperature for $UTe_2$ with H along the *b* axis. Phase boundaries have been defined by Refs. *[4]* (circles), *[59]* (downward triangles), *[58]* (squares), *[5]* (upward triangles), and *[60]* (diamonds). The line indicating the transition into the field polarized state is based on data from Refs. [58-60].

### 3.2 Magnetic transition

When an external magnetic field is applied along the *b* axis of $UTe_2$, the system undergoes a magnetic transition into a field-polarized (FP) state at 35 T as spins—and the easy magnetization axis—abruptly rotate from the *a* axis to the *b* axis [12, 14]. This is often called the metamagnetic transition, referring to the sudden rise in the magnetization with a small change in applied field, as shown in Figure 6(a). The magnetic moment along the *b* axis jumps dramatically at $H_m$, the critical field of the transition. The magnitude of the jump has been reported as 0.6 $\mu_B$ per formula unit [14] and as 0.3 $\mu_B$ per formula unit [12]; the discrepancy may be due to misalignments in applied magnetic field or to sample dependence.

For magnetic fields in the crystalline *bc* plane, the metamagnetic transition seems to occur when the component of field along the *b* axis is roughly 35 T, regardless of the magnitude of field along the *c* axis [12]. In other words, $H_m$ may be proportional to $1/\cos(\theta)$, where $\theta$ is the angle from the *b* to *c* axes. If so, then in theory the metamagnetic transition should occur even for $\theta$ of nearly 90 degrees, as long as a high enough field could be applied. The transition field $H_m$ rises much more rapidly in the *ab* plane than it does in the *bc* plane as the magnetic field is tilted away from the *b* axis, as can be seen in Fig. Figure *3*. This can be understood given that the metamagnetic transition is a flip of spins from the *a* axis to the *b* axis. In the *bc* plane, only the component of field along *b* matters; but in the *ab* plane, any tilt away from the *b* axis is also a reinforcement of moments along the *a* axis and works against the metamagnetic transition.

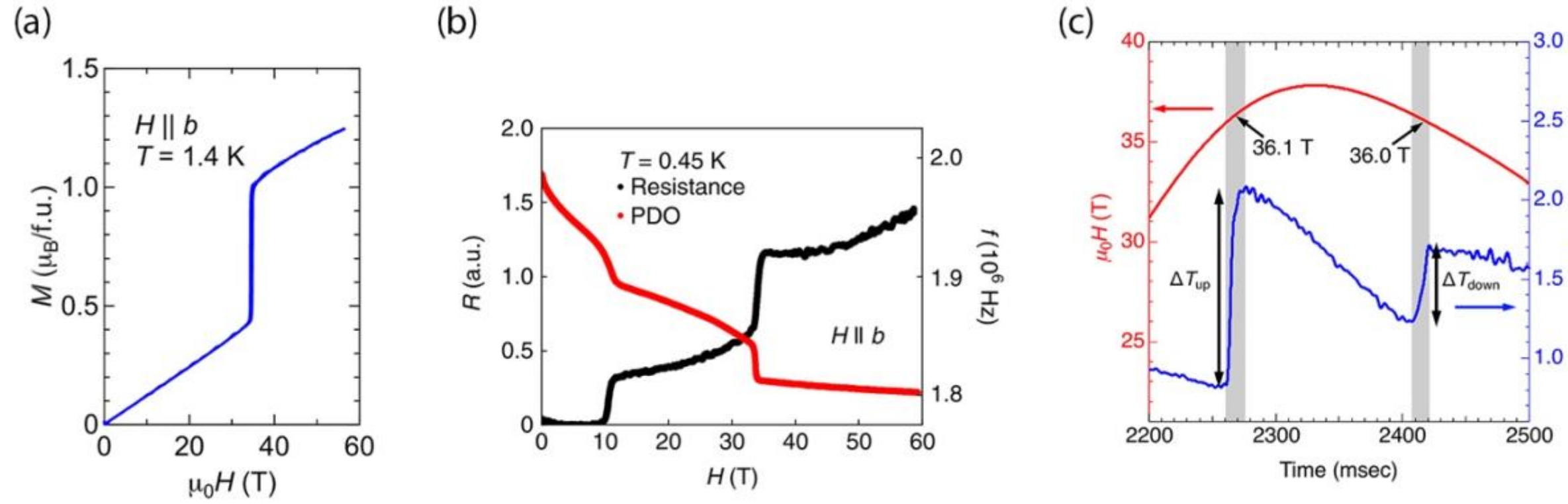

Figure 6. Various signatures of the metamagnetic transition in $UTe_2$ for H ∥ b axis: (a) magnetization, from Ref. *[14]*; (b) magnetoresistance and PDO signal (slightly off *b* axis), from Ref. *[12]*; and (c) magnetocaloric effect, from Ref. *[67]*. All show a sharp discontinuity at the transition.

The first-order nature of the phase transition at low temperatures is indicated by the presence of hysteresis loops in magnetization versus field and resistance versus field [14, 15]. While the value of $H_m$ changes very little with temperature, the difference in $H_m$ for rising versus falling fields does shrink as the temperature increases. For field along the *b* axis, the hysteresis loop vanishes at roughly 11 K [14] or 7 K [15], with $\mu_0 H_m \approx 34.7$ T; the difference in reported temperatures is likely due to sample variation. This temperature is therefore identified as a critical end point, with metamagnetic transitions above this temperature being second-order. Similarly, it has been reported that the Seebeck coefficient only shows hysteresis at $H_m$ below approximately 7 K [59].

At the metamagnetic transition, there is a jump in the magnetoresistance. When the longitudinal resistivity is fit to the Fermi-liquid form $\rho = \rho_0 + AT^2$, a distinct peak in the coefficient *A* is noted at $H_m$; this suggests an enhancement of the effective mass, perhaps due to critical magnetic fluctuations [15]. Another signature of the transition is a sudden decrease of measured frequency when the sample is measured using a proximity detector oscillator (PDO) circuit, as described in Ref. [68]. Since PDO is sensitive to many sample properties, primarily electrical conductivity and magnetic permeability, it is difficult to assign the change in PDO frequency to a specific physical mechanism; that said, the sudden decrease in frequency clearly demarcates the phase transition, as shown in Figure 6(b). The Seebeck coefficient of $UTe_2$ also undergoes a sudden jump at $H_m$, though the magnitude and even sign of the jump are temperature-dependent [59]. Magnetostriction measurements show a step-like decrease in the unit cell volume across $H_m$; across the metamagnetic transition, the *a*-axis lattice parameter shrinks proportionally to the unit cell volume but the magnitudes of *b*- and *c*-axis magnetostriction are larger and are opposite to each other in sign, suggesting a *bc* plane lattice instability [69].

It has been hypothesized that these abrupt changes in physical properties are due to a Fermi surface reconstruction at $H_m$ [59]. The Fermi surface of $UTe_2$ in its field-polarized state has not yet been determined experimentally, and therefore cannot be compared to its zero-field Fermi surface; direct measurement of the high-field Fermi surface through quantum oscillations would be of great interest. In any case, whether or not there is a topological discontinuity such as a Fermi surface reconstruction at $H_m$, there is clearly a sudden change in the electronic structure of $UTe_2$.

The magnetocaloric effect of a single crystal of $UTe_2$ has been measured across the metamagnetic transition at low temperature, as shown in Figure 6(c) [67]. These measurements were taken with H aligned a few degrees from the *b* axis. The experimental results are consistent with a first-order transition. A detailed thermodynamic analysis indicates that the change in entropy at the metamagnetic transition, as well as the sample's hysteretic thermal loss, are predominantly due to the change in magnetization of $UTe_2$ across the transition [67].

The effect of the metamagnetic transition on the heat capacity of $UTe_2$ has also been studied. The heat capacity of a sample as a function of temperature was measured at various fixed magnetic fields and the heat capacity in the normal state was modeled as $C_p = \gamma T + \beta T^3$. Based on this fitting, it was found that there is a peak in γ, the Sommerfeld coefficient in the specific heat, at the metamagnetic transition with field near the *b* axis [67]. The peak in γ near $H_m$ was also predicted from magnetization data, using a thermodynamic Maxwell relation [67]. The relevant heat capacity data and magnetization data are shown in Figure 7(a) and Figure 7(c), respectively. The extracted values of γ, divided by its zero-field value, are shown in Figure 7(b); there is roughly a twofold enhancement of γ near the metamagnetic transition, with field along the *b* axis. Due to the change in entropy at the metamagnetic transition, there should also be a discontinuous step down in γ at $H_m$; while this detail cannot be resolved in the data, it is consistent with the observation that γ is not symmetric about $H_m$ and is lower in the high-field state [67]. For H along the *b* axis, a small discontinuous drop in γ just above $H_m$ was demonstrated indirectly, through a thermodynamic analysis employing the magnetic Clausius-Clapeyron equation [70].

For a Fermi liquid, γ is proportional to the quasiparticle effective mass; therefore, the data suggest that there is a peak in effective mass at $H_m$. When electrical resistivity versus temperature is fit with the form $\rho_0 + AT^2$, then *A* should also be proportional to the effective mass squared in a Fermi liquid. Knafo *et al.* fit values of *A* for various fields along the *b* axis and found that it too is enhanced near $H_m$ [15]. Figure 7(b) shows the square root of *A* divided by its zero-field value, plotted alongside the extracted values of γ discussed above. If a Fermi liquid model is appropriate, then it appears effective mass is roughly doubled near $H_m$.

As pointed out in Ref. [14], if it is assumed that the band structure of $UTe_2$ does *not* change with field, then the enhancement of the effective mass near $H_m$ must be driven by correlations—presumably ferromagnetic fluctuations. The authors hypothesize that this increase in effective mass is directly responsible for the enhancement of superconductivity on approaching $H_m$ when magnetic field is near the *b* axis. An extensive analysis of the assumptions underlying this hypothesis can be found in Sec. 6.5.1 of Ref. [13]. The enhancement of effective mass is directly related to the electron-phonon coupling constant for strongly coupled superconductors [71]. However, for a system such as $UTe_2$ in which the superconductivity is likely driven by magnetic fluctuations rather than electron-phonon pairing such a direct connection is not evident. Calculated properties of $UGe_2$ based on magnetically mediated superconductivity suggest that effective mass in this material is also enhanced near its metamagnetic transition, connected to strengthened superconductivity at that point in the phase diagram [72, 73]. Yet this is in the context of a ferromagnetically ordered superconductor, whereas in $UTe_2$ there is no long-range magnetic order at ambient pressure. Rather, for $UTe_2$ it may be the case that the mass renormalization and the superconducting pairing each have their own magnetic field dependence [74]. As seen in Figure 3, enhanced superconductivity only appears for a narrow angular range of magnetic field direction near the *b* axis, while the metamagnetic transition—and the accompanying enhancement of effective mass—occurs for a much broader angular range.

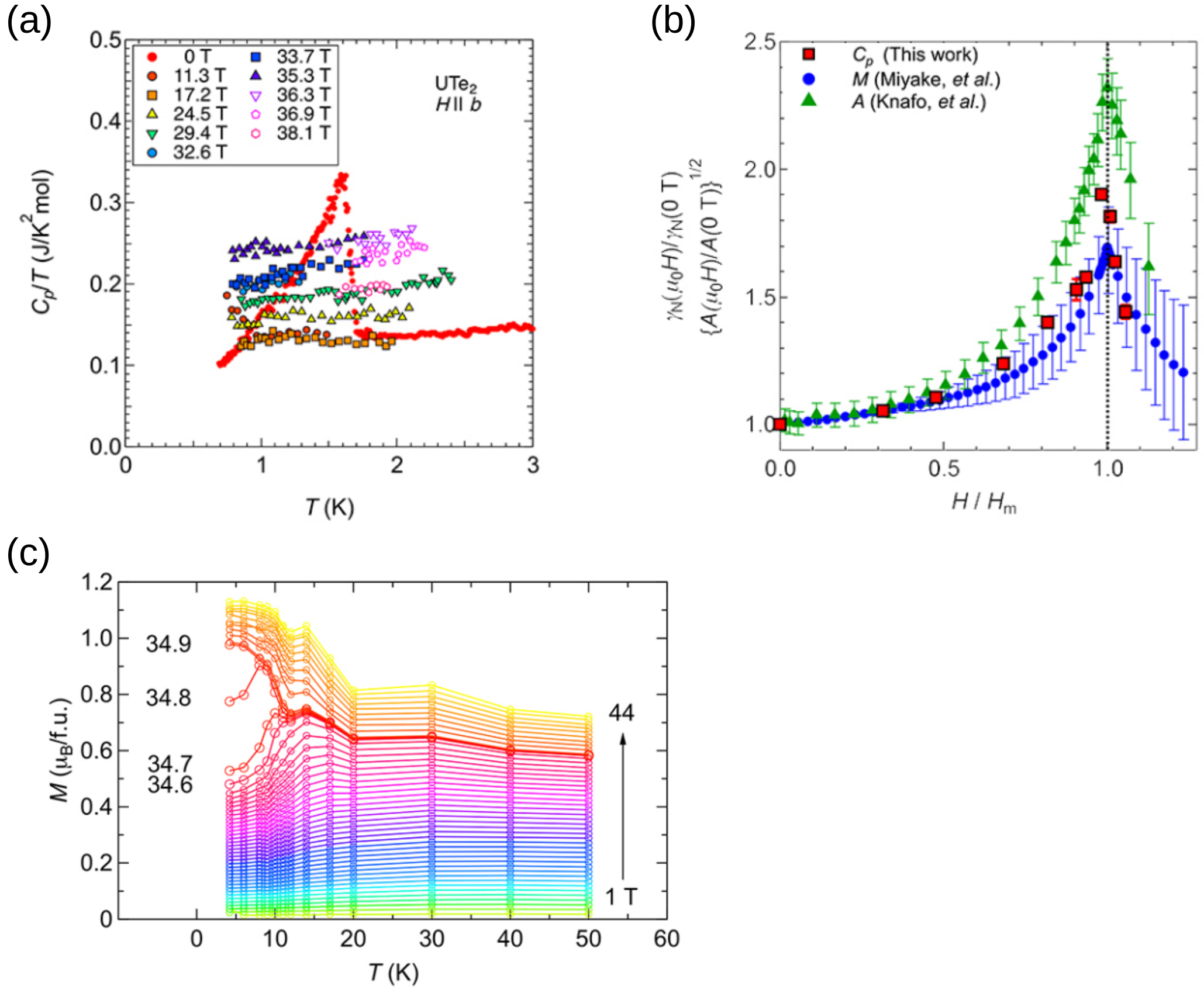

Figure 7. (a) High-field heat capacity with field along the *b* axis, from Ref. [67]. (b) Enhancement of γ and of *A* (the quadratic coefficient of the electrical resistivity) for fields near $H_m$ along the *b* axis, from Ref. [67]. (c) The temperature dependence of the magnetization with constant fields along the *b* axis, from Ref. [14]. The red squares and blue circles in subfigure (b) show γ/γ(0 T), where red squares represent γ found from extrapolation of the heat capacity data in subfigure (a) to zero temperature and the blue circles show γ found by analyzing the magnetization data in subfigure (c) using a thermodynamic Maxwell relation. The green triangles show the square root of *A/A*(0 T), where *A* was found from fits to electrical resistivity with field along the *b* axis in Ref. [15].

The metamagnetic transition persists to higher temperatures than any of the observed superconducting phases of $UTe_2$. At low temperatures there is a sharp jump in magnetization at $H_m$. With increasing temperature, the jump becomes less sharp and lower-magnitude, but there is still an obvious change in slope of the magnetization versus temperature up to roughly 20 K [14]. Similarly, the peak in resistance associated with the metamagnetic transition broadens greatly with increasing temperature, but can be distinguished up to about 25 K [60].

This is similar to the extrapolated zero-field temperature at which there is a maximum in *b*-axis magnetic susceptibility, denoted $T_\chi^{max}$ [4, 14]. It has been suggested that these features are related and that in addition to $H_m$ marking a magnetic phase transition, $T_\chi^{max}$ and the maxima of resistivity may mark the approximate boundary of a correlated paramagnetic regime [4, 14, 60].

As a related hypothesis, Willa *et al.* suggest a connection between $H_m$ and a Schottky-like anomaly that appears at approximately 12 K in zero field, and attribute $T_\chi^{max}$ to the tail of this anomaly [75]. In addition to the maximum in *b*-axis susceptibility, there is also a peak in the Hall coefficient at a similar temperature to $T_\chi^{max}$, evolving similarly with field [59]. Notably, $T_\chi^{max}$ is similar to the Kondo lattice coherence temperature of $UTe_2$. It may be the case that the termination of a correlated paramagnetic regime, both at $T_\chi^{max}$ and at $H_m$, is related to the waning of Kondo hybridization. Inelastic neutron scattering has revealed magnetic excitations that appear to be due to Kondo hybridization, which are only discernable below the coherence temperature [56]. It would be interesting to determine whether these magnetic excitations also disappear at $H_m$; if so, it would solidify both the connection between $T_\chi^{max}$ and $H_m$ as well as the role of Kondo interactions in the phase diagram of $UTe_2$.

### 3.3 Field-polarized superconducting phase

While $H_m$ marks the upper field boundary of the SC2 phase, it is also the lower field boundary for a field-induced superconducting phase, as shown in the resistance data in Figure 4(b). This superconducting phase of $UTe_2$ only exists within the field-polarized (i.e., polarized paramagnetic) regime and is thus referred to as the $SC_{FP}$ phase.

The $SC_{FP}$ phase has been studied for magnetic fields in the *bc* plane; in this plane, it arises for θ between roughly 20 degrees to 40 degrees [12]. As noted in Ref. [60], this is near the (0 1 1) direction in reciprocal space at $\theta \approx 23.7°$; whether this has a relationship to the superconducting pairing is unknown. This is another instance in which future studies with precise crystal alignments may be helpful. The field and temperature evolution of the $SC_{FP}$ phase for three different magnetic field angles in the *bc* plane is shown in Figure 8. A recent preprint reports magnetocaloric measurements that indicate $SC_{FP}$ is indeed a bulk superconducting phase [61]. A separate preprint reports that the normal-state Hall effect in the FP state is related to the $SC_{FP}$ phase, with the Hall effect almost entirely suppressed where the $SC_{FP}$ phase is most robust [76]. The authors hypothesize that this could be due to compensation from an internal exchange field, as required for the Jaccarino-Peter mechanism described in Sec. 3.5.5.

The same thermodynamic analysis that showed an expected drop of γ above $H_m$ for H along the *b* axis also indicates a small discontinuous jump in γ just below $H_m$ with $\theta \approx 28°$ [70]. In other words, γ drops discontinuously as the SC2 phase is exited and rises discontinuously as the $SC_{FP}$ emerges, indicating a possible connection between the effective mass enhancement and the superconducting pairing. As yet, the superconducting pairing mechanism involved in the $SC_{FP}$ state has not been determined.

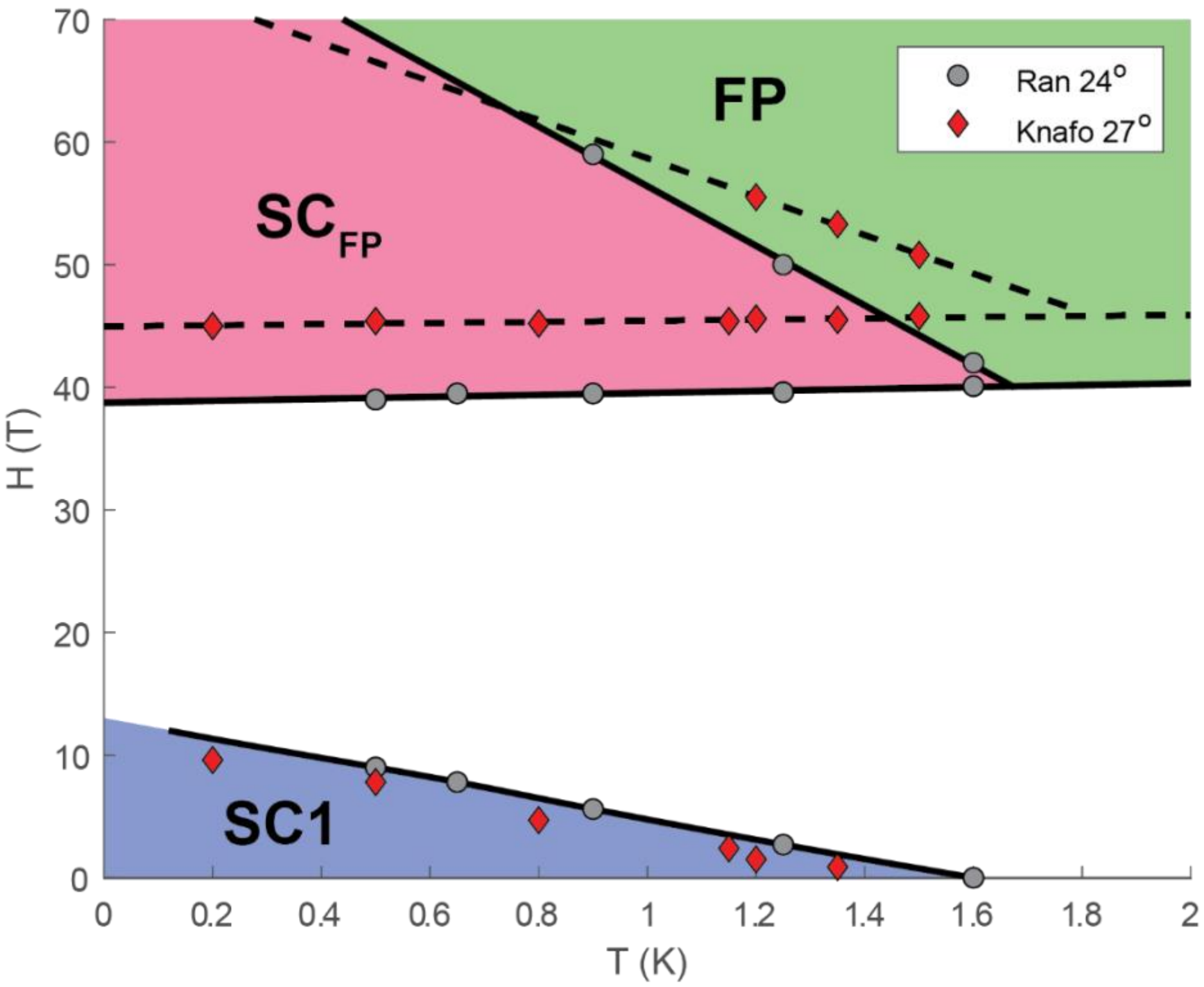

Figure 8. Phase diagrams for $UTe_2$ as a function of temperature and magnetic field strength, for two similar angles of field within the *bc* plane. Data at 24° from the *b* axis are from Ref. *[12]*; data at 27° from the *b* axis are from Ref. *[60]*.

### 3.4 Field along *a* axis

With field directly along the *a* axis of $UTe_2$, there is no metamagnetic transition or other obvious phase transition in $UTe_2$. It has been proposed that there is a field-induced Lifshitz transition when roughly 5.6 T is applied along the *a* axis, based on subtle features in transport [77]. As a function of field, there is a peak in the magnitude of thermoelectric power at 5.6 T, accompanied by a change in slope of the thermal conductivity. At temperatures approaching 0 K, this field is close to the $H_{c2}$ of $UTe_2$ for fields along the *a* axis. However, while $H_{c2}$ rapidly decreases with increasing temperature, the feature in thermoelectric power and thermal conductivity appears to be temperature independent.

The thermoelectric power as a function of temperature has also been studied with various fixed field strengths along the *a* axis. With field fixed at 5.6 T, there is a narrow temperature region above $T_c$ in which the thermoelectric power divided by temperature goes as $T^{-1/2}$ [77]. A $T^{-1/2}$ dependence is expected near a neck-disruption type Lifshitz transition [78]. There should also be a subtle kink in conductivity as a function of field if this indeed a neck-breaking Lifshitz transition, and the longitudinal conductivity near 5.6 T should go roughly as $T^{1/2}$ [78, 79]. Measurements of both thermoelectric power and electrical conductivity on the same sample as functions of field and temperature would be useful to clarify whether this is indeed a Lifshitz transition. If so, the question then would arise as to whether the post-transition Fermi surface in this case has any relation to the Fermi surface in the field-polarized state. As pointed out in Ref.

[77], this putative Lifshitz transition occurs at roughly the same value of magnetization along the *a* axis as the value of magnetization along the *b* axis for the metamagnetic transition.

### 3.5 Proposed explanations for high-field superconducting phases

In addition to the enhancement of spin fluctuations discussed above, several different theories have been proposed to explain the various superconducting phases of $UTe_2$. Below, we will lay out the basic assumptions and mechanisms of the prevailing theories. See the review by Aoki *et al.* for a detailed analysis of the type and position of gap nodes associated with the order parameters discussed in this section [13]. Of note, some of these theories may be complementary; one may prove to describe the SC2 phase while another explains the $SC_{FP}$ phase.

#### 3.5.1 Free rotation of **d** vector, broken time-reversal symmetry

One theoretical explanation for the reinforcement of superconductivity with field along the *b* axis requires that the triplet state is nonunitary, meaning that it must break time-reversal symmetry [80]. This assumption has support from experiments: measurements of a spontaneous polar Kerr effect in $UTe_2$ are consistent with broken time-reversal symmetry in the superconducting state [10]. In this case, the order parameter can couple to the magnetization in free energy such that when the magnetization changes, the **d** vector that describes the order parameter will rotate. In this model, there is essentially a rotation of the **d** vector as field is increased to go from the SC1 state to the SC2 state. This model can reproduce the temperature dependence of the *b*-axis $H_{c2}$. Additionally, it is consistent with the evolution of the specific heat of $UTe_2$ as a function of temperature and field angle [9]. This model also can explain the $SC_{FP}$ phase, as it yields a superconducting pocket at high fields that appears for a narrow range of magnetic field directions near $\theta = 35°$ [80]. Of note, this model relies on the assumption that the effective many-body spin-orbit coupling is weak enough that the **d** vector is not locked to the lattice structure.

#### 3.5.2 Constrained change of **d** vector, broken time-reversal symmetry

If strong spin-orbit coupling is assumed, in contrast to the above proposal, then the symmetry of the superconducting order parameter is constrained to the $D_{2h}$ point group symmetry of $UTe_2$. Under this symmetry, the SC1 state in zero field can only break time-reversal symmetry if it has a two-component order parameter with components from two different irreducible representations. The possibility of SC1 having a two-component, time-reversal symmetry breaking order parameter, while SC2 has only a single-component order parameter was first examined in the Supplementary Information of Ref. [16].

If, beyond breaking time-reversal symmetry, the SC1 order parameters can couple to a *c*-axis magnetic field as suggested in Ref. [10], then the only possible combinations of order parameters are those shown in Table 1. If the superconductivity of $UTe_2$ is spin-triplet, the possible combinations of order parameters for the SC1 state are further limited to only (i) and (ii) shown in Table 1 [10].

| | Irrep of $\psi_1$ | Irrep of $\psi_2$ |
|---|---|---|
| i | $B_{3u}$ | $B_{2u}$ |
| ii | $B_{1u}$ | $A_u$ |
| iii | $B_{3g}$ | $B_{2g}$ |
| iv | $B_{1g}$ | $A_g$ |

Table 1. Allowed irreducible representations of the SC1 order parameters of $UTe_2$, given broken time-reversal symmetry that couples to *c*-axis magnetic field and assuming strong spin-orbit coupling; from Ref. [10].

This should result in two distinct superconducting transitions as a function of temperature; as mentioned in the previous section, certain samples of $UTe_2$ consistently show two transitions in specific heat, but samples grown in slightly different growth conditions show only a single transition. If this model is correct, it would have to be the case that samples with an observed single transition actually have two transitions that are so near in temperature as to be functionally indistinguishable.

One microscopic model consistent with these symmetry constraints is laid out by Shishidou *et al.* [81]. Their density functional theory plus Hubbard $U$ (DFT + $U$) calculations identify a topological band near the chemical potential of $UTe_2$ for all values of $U$ that were studied. These calculations also indicate the importance of the rungs of uranium atoms that form a ladder-like structure in the crystal; it is only by considering the sublattice rung degree of freedom that this topological band can arise from a simple model Hamiltonian [81]. Based on their model and the polar Kerr effect measurements, they find that the superconducting pairing should be $B_{3u} + iB_{2u}$ below the second transition. Their predictions for the evolution of the transitions with magnetic field and temperature are shown in Figure 9. For field along the *b* axis, the $B_{2u}$ pairing state should experience much less paramagnetic limiting than the $B_{3u}$ pairing does. This naturally would lead to the existence—and persistence in *b*-axis field—of an SC2 state with $B_{2u}$ order parameter. This model does not address the $SC_{FP}$ phase.

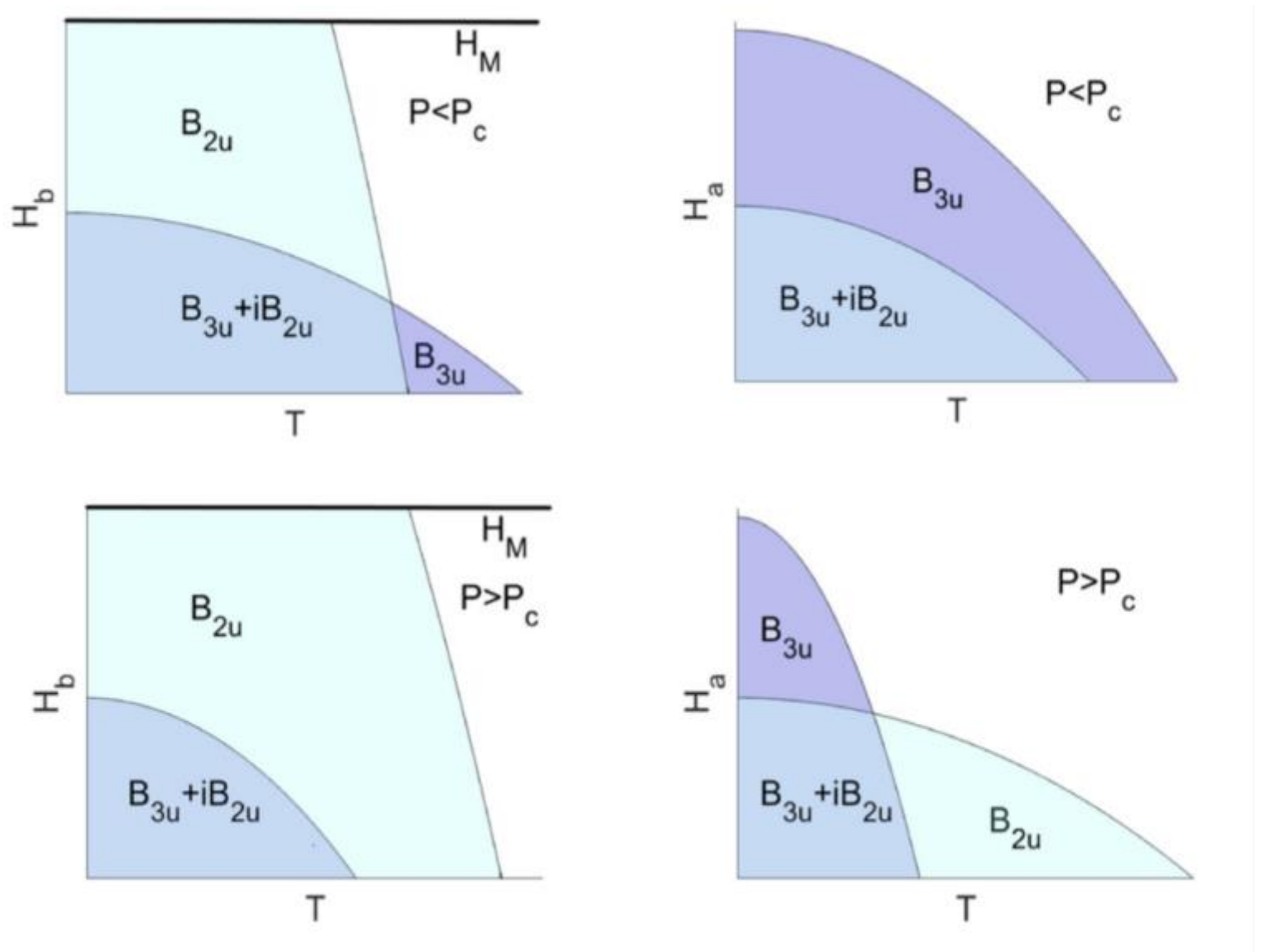

Figure 9. Qualitative phase diagrams of $UTe_2$ for fields $H_a$ along the *a* axis and $H_b$ along the *b* axis, from Ref. *[81]*. The upper (lower) two diagrams are for pressures below (above) approximately 0.2 GPa.

3.5.3 Constrained change of **d** vector, preserved time-reversal symmetry

If one is skeptical of the experimental evidence for broken time-reversal symmetry, then models with preserved time-reversal symmetry can also be constructed. As long as strong spin-orbit coupling is assumed, the order parameter will still be limited to irreducible representations of the point group $D_{2h}$. However, without the constraint of broken time-reversal symmetry, the order parameter can be any (or any combination of) the eight irreducible representations of $D_{2h}$. Determining the appropriate order parameter(s) then relies on microscopic models, which can be informed by experimental evidence for the gap structure of the superconducting phases.

One such model is the 24-band tight-binding model of Ishizuka *et al.* [82]. This periodic Anderson model reasonably reproduces the Fermi surfaces seen in angle-resolved photoemission spectroscopy, as well as the low-energy band structure of $UTe_2$ found through DFT + *U* calculations. Based on this model Hamiltonian, eigenvalues of the Eliashberg equation can be found for the various $D_{2h}$ irreducible representations. The eigenvalue solutions indicate that the superconducting order parameter at ambient pressure should belong to $B_{3u}$ or $A_u$, an odd-parity spin-triplet superconducting state that preserves time-reversal symmetry. From the linearized Eliashberg equation, it was also determined that the nearly-degenerate $B_{3u}$ and $A_u$ superconducting states in this model should yield a **d** vector that is mostly in the crystallographic *bc* plane. In other words, the spin of the Cooper pairs is mainly along the *a* axis. This would lead to an enhanced upper critical field with field along the *a* axis, as paramagnetic limiting does not affect spin-triplet superconductors when the field and spin are parallel. However, the periodic Anderson model does not seem to explain the SC2 state, i.e. enhanced superconductivity along the *b* axis of $UTe_2$. This model also does not address the $SC_{FP}$ state.

3.5.4 Confinement to two dimensions

In any quasi-two-dimensional or layered quasi-one-dimensional superconductor, a magnetic field parallel to the conducting layers should first suppress and then reinforce or restore superconductivity as field is increased [83]. In this vein, simple models of $UTe_2$ with triplet pairing and cylindrical Fermi pockets are able to qualitatively reproduce the observed behavior of $T_c$ as a function of field [16, 84, 85]. As described above, ARPES and recent quantum oscillation measurements do indicate that $UTe_2$ has cylindrical Fermi pockets along $k_z$.

Lebed gives a formula to estimate $H^*$, the magnitude of field along the $b$ axis of $UTe_2$ necessary to cause the two-dimensional confinement that would reinforce superconductivity [84]. Based on tight-binding models and the generalized Helfand-Werthamer framework of the orbital upper critical field, along with the experimental results of Ref. [9], the authors of Ref. [13] conclude that $H^*$ is over 1000 T. However, it must be noted that even for field along the relatively conventional $c$ axis, the temperature-dependence of $H_{c2}$ for $UTe_2$ does not follow the predictions of the Werthamer- Helfand-Hohenberg model [4]. Though the estimated $H^*$ suggests its implausibility, the uncertainties of both the Fermi surface of $UTe_2$ and its underlying physics make it difficult to entirely rule out this mechanism of reinforced superconductivity as an explanation for the SC2 or $SC_{FP}$ states.

### 3.5.5 The Jaccarino-Peter mechanism

In some magnetic materials, the conduction electrons experience an internal exchange field that opposes an applied external magnetic field. In this case, once the applied field is strong enough to cancel out the effective internal magnetic field, there is a restored degeneracy between opposite-spin electrons that allows the formation of a singlet superconducting state [86]. This phenomenon, called the Jaccarino-Peter effect, can explain field-induced superconductivity in several Chevrel phase materials and organic superconductors [87-89].

Notably, such a mechanism cannot compensate the Meissner current, assuming that the magnetic ions involved are well localized and provide short ranged exchange interactions [90]. In other words, the Jaccarino-Peter mechanism describes effective field compensation for the *paramagnetic* pair-breaking effect, but such compensation will not occur for *orbital* pair-breaking effects.

Paramagnetic limiting may occur in spin-triplet superconductors as long as there is a non-zero component of the **d** vector along the direction of the applied field; see Ref. [64] for a detailed explanation. Therefore, the Jaccarino-Peter effect can arise in spin-triplet superconductors as well. However, given the above-described NMR evidence that there is no $b$-axis component of the **d** vector in the SC2 phase, it is more likely that paramagnetic limiting is simply not extant for fields along the $b$ axis of $UTe_2$, rather than the more complicated explanation of paramagnetic limiting existing but being compensated by an internal $b$-axis exchange field.

It is not yet known whether the $SC_{FP}$ phase is spin-singlet or spin-triplet in nature, but in either case the Jaccarino-Peter effect could conceivably be relevant. However, if the Jaccarino-Peter effect were applicable to either the SC2 or $SC_{FP}$ phase of $UTe_2$, some further mechanism would have to be at play to account for suppression of orbital pair-breaking effects.

### 3.5.6 Superconductivity near the quantum limit

One proposed mechanism for field-enhanced superconductivity in general is the quantization of conduction electrons into Landau levels [91]. In order for Landau levels to be relevant and not smeared by thermal scattering, it must be the case that $k_B T \ll \hbar\omega_c$, where $\omega_c = \frac{eB}{m_c}$ is the cyclotron frequency. The cyclotron mass, $m_c$, is determined by the specific orbit of a quasiparticle around the Fermi surface, not simply by band structure. We do not have a reliable determination of $m_c$ for $UTe_2$ with field along the $b$ axis or with field along the special angle in the $bc$ plane that leads to the $SC_{FP}$ phase. However, recent measurements of quantum oscillations indicate that for $UTe_2$ quasiparticles, the cyclotron masses for magnetic field in the $ac$ plane range between $32m_0$ and $57m_0$ [39]. The transition temperature of the $SC_{FP}$ phase of $UTe_2$ is approximately 1.6 K [12]. At this temperature, we find limits of $38.1\,T \ll B$ for $32m_0$ and $67.9\,T \ll B$ for $57m_0$. Given this range, it is conceivable that we are in the Landau level regime of $UTe_2$ for the $SC_{FP}$ phase. A theoretical work has proposed that the $SC_{FP}$ phase in $UTe_2$ may be due to Landau quantization in a 'Hofstadter butterfly' regime with large superlattices [92]. As the authors of that study note, in this scenario we would expect there to be an even higher-field superconducting region in $UTe_2$ where the next Landau level crosses the Fermi surface. This scenario is of possible relevance to behavior observed at high pressure, as discussed in Sec. 4.3.2.

## 4. Evolution under pressure

Compared to substitutional or other induced disorder-based tuning, pressure is a powerful, clean method with which to affect the electronic structure of unconventional superconductors [93, 94]. This is especially important in the case of $UTe_2$, as the strong fluctuations observed in muon spin relaxation/rotation (μSR) experiments indicated that superconductivity in this material lies near a ferromagnetic quantum critical point [48], suggesting that a quantum phase transition could be achievable with pressure. In practice, $UTe_2$ is indeed very sensitive to the application of hydrostatic pressure, even at zero field. Once magnetic fields are applied, the resulting phase diagrams for superconductivity and magnetism are quite complex. We will mainly focus on the pressure-field phase diagrams of $UTe_2$ with temperature as close to 0 K as possible. For an overview of the evolution of these phase diagrams with increasing temperature, for fields along the $b$ and $c$ axes, see the electrical resistivity studies of Vališka *et al*. [18].

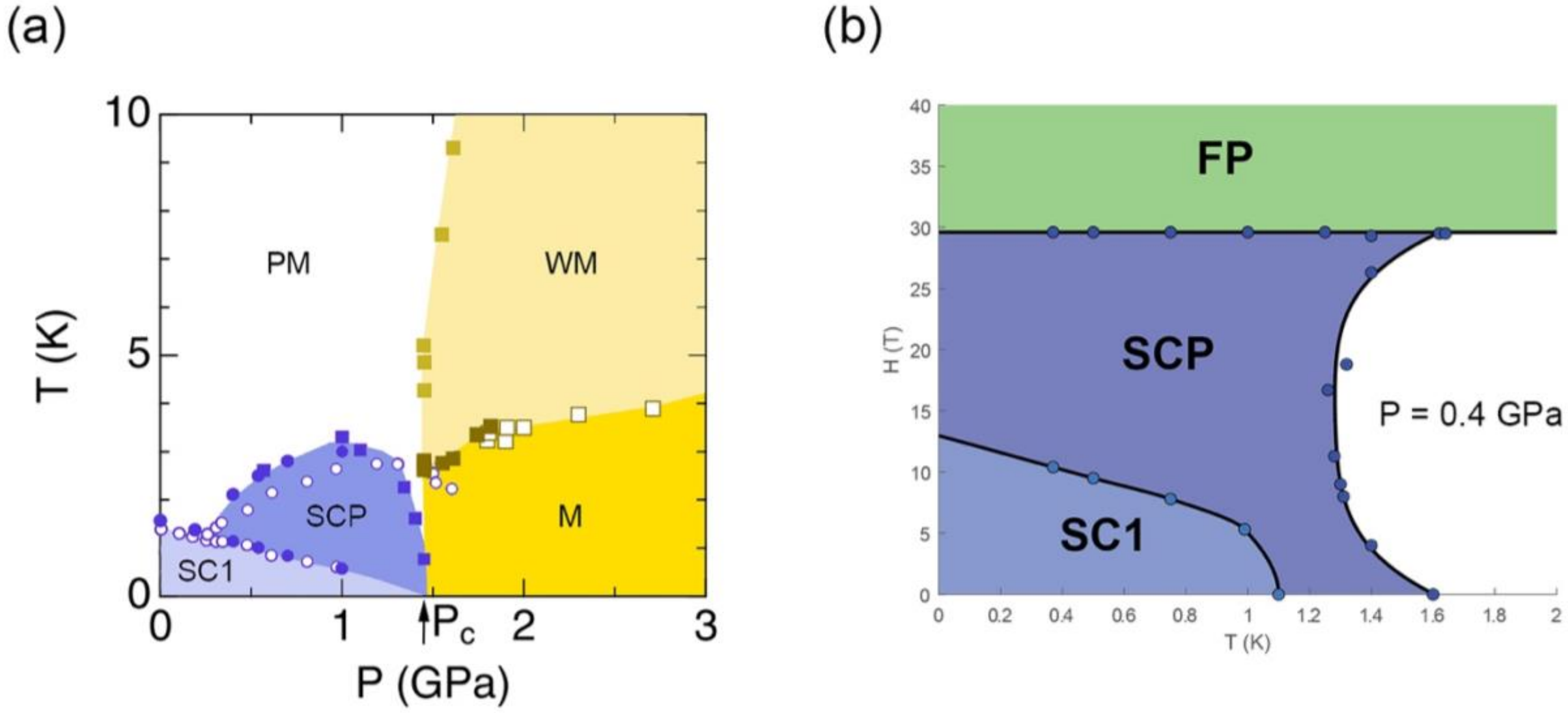

Figure 10. (a) The pressure-temperature phase diagram of $UTe_2$ in zero magnetic field, adapted from Ref. [20]. The open circles and open squares are from specific heat data from Ref. [95]; solid circles are from the magnetoresistance and AC calorimetry measurements of Ref. [96]; solid squares are from the magnetoresistance measurements of Ref. [20]. The SC1 and SCP phases, as well as the M and WM phases, are described in the text; PM stands for paramagnetic. (b) The temperature-field phase diagram

of $UTe_2$ at 0.4 GPa, with field along the b axis. The phase boundaries are from Ref. [16], based on resistance and tunnel diode oscillator data.

### 4.1 Behavior at low pressures

Figure 10(a) shows our current understanding of the pressure-temperature phase diagram of UTe2 in zero field. As pressure is initially applied, (0.1-0.3 GPa) the ambient pressure zero-field transition shifts down slightly in temperature (approximate reduction of 0.5 K by 0.3 GPa) [95]. As mentioned in Section 2**,** there is some debate about whether this ambient-pressure transition can reliably be considered a single or double transition. However, above 0.3 GPa, multiple low-field transitions are universally acknowledged [19, 95-97]. We will refer to the superconducting phase that appears under pressure as SCP. The SCP transition appears at higher temperature than the SC1 transition and increases with pressure to a maximum at ~1 GPa before rapidly decreasing towards a critical pressure $P_c$ (~1.5 GPa), whereas the lower temperature transition decreases linearly with pressure, extrapolating to 0 K at $P_c$ as shown in Figure 10(a). The highest superconducting transition observed for $UTe_2$ in any investigation thus far was recorded by Ran *et al.* at 1 GPa, with a full twofold increase in resistive $T_c$ at this pressure ($T_{cmax}$ = 3.2 K) [17]. According to subsequent magnetization measurements under pressure, superconductivity at $T_{cmax}$ is indeed bulk [17]. There is some disagreement about the exact critical pressure ($P_c \gtrsim 1.4 - 1.7$ GPa) above which low-field superconductivity is completely suppressed. These differences, as well as slight variations in maximum $T_c$ with pressure, are likely attributable to pressure step-size differences compounding sample variation issues.

### 4.2 Behavior at or above $P_c$

NMR measurements of $UTe_2$ indicate that there is a discontinuous change in its electronic structure at $P_c$, likely due to destruction of the Kondo coherent state [98]. Above $P_c$ the heat capacity of $UTe_2$ exhibits two local maxima as a function of temperature, indicating phase transitions [97], which correspond to anomalies in normal-state resistivity [17, 18, 20, 95, 97]. Based on the broad nature of the features at $T_{m1}$ (≈ 4 K), it has been posited that the higher-temperature transition represents the development of short-range weakly magnetic (WM) order while the lower temperature $T_{m2}$ (≈ 3 K) is the onset of long-range magnetic (M) order [17, 20, 95, 97]. $T_{m2}$ is largely unaffected by pressure until about 3 GPa [17, 18, 20, 95, 97], above which the temperature of the transition rises moderately [95]. The temperature of $T_{m1}$, however, increases sharply with pressure, as shown in Figure 10(a) [17, 20, 97].

### 4.3 High-field behavior with applied pressure

The upper critical field of $UTe_2$ is highly anisotropic; at ambient pressure, $H_{c2}^{b} > H_{c2}^{c} > H_{c2}^{a}$. However, Knebel *et al.* note that on approaching $P_c$ the magnetic anisotropy changes significantly, and $H_{c2}^{c} > H_{c2}^{a} \sim H_{c2}^{b}$ [19]. By analogy with other heavy fermion superconductors, in which the anisotropy of $H_{c2}$ is tied to the anisotropy of susceptibility, Knebel *et al.* suggested that the magnetic hard axis may shift from *b* to *c* near $P_c$.

This hypothesis was shortly confirmed by magnetization measurements under pressure. As $UTe_2$ is brought from ambient pressure to 1.7 GPa, the easy magnetic axis for low temperatures changes from the *a* axis to the *b* axis, while the hard magnetic axis for low temperatures changes from the *b* axis to the *c* axis [99]. The metamagnetic transition also involves a change of the easy magnetic axis from *a* to *b*, an indication that the field-polarized state existing at high pressures is the same state that is seen above the metamagnetic transition. This connection may be more apparent by studying the pressure-field phase diagrams of $UTe_2$, such as those shown in Figure 11; we can see that there is a continuous evolution from some of the high-pressure phases at low field to some of the ambient pressure phases at high field.

The relationships between the high-field and high-pressure phases of $UTe_2$ can help us to understand the underlying physics of both. Below, we will describe the behavior of $UTe_2$ under pressure when fields are applied along the three crystallographic axes, and along a direction in the *bc* plane that induces the $SC_{FP}$ phase at ambient pressure.

#### 4.3.1 Field along *b*

Figure 10(b) shows a phase diagram as a function of temperature and *b*-axis field when $UTe_2$ is held at a constant pressure of roughly 0.4 GPa. The boundary between the SC1 and SCP phases is derived from tunnel diode oscillator (TDO) measurements, based on observed kinks in TDO frequency versus field [16]. Similar TDO measurements at several pressures, with temperature held at approximately 0.4 K, were used to create the pressure-field phase boundaries of the SC1 phase shown in Figure 11(a). The ambient pressure data point was extracted from Ref. [12] by extrapolating the SC1 critical field to field direction along the *b*-axis; this inherently assumes that SC1 and SC2 are distinct phases. If so, there is an apparent evolution from the SC1-to-SC2 boundary to the SC1-to-SCP boundary as pressure is applied.

Calorimetry measurements in both applied pressure and applied *b*-axis field would be helpful to determine whether there is a pressure-induced phase transition between the SC2 and SCP phases, or whether the two phases are one and the same. The superconducting phase that is stabilized by pressure at zero field may be the same SC2 phase that emerges at ambient pressure with an application of field along the *b* axis of $UTe_2$.

The evolution of the metamagnetic transition has also been measured through magnetoresistance and magnetic susceptibility measurements, as well as TDO experiments [16, 18, 19]. As seen in Figure 11(a), when H is || *b*, the field of the metamagnetic transition is suppressed monotonically by pressure, as is the boundary between the SC1 and SCP phases.

Above $P_c$ there is a broad maximum in resistivity versus temperature, which is thought to indicate a transition into an ordered magnetic state; this feature remains until about 15 T, when the ordered magnetic state is truncated by the transition into the field-polarized regime [16]. Lin *et al.* posit that the transition from magnetic order to a field-polarized state is first-order due to hysteresis observed in magnetoresistance at 1.88 GPa [16].

It is worth noting that the field-polarized phase evolves continuously from ambient pressure, at which it only exists with high fields applied near the *b* axis, to a state that is stable at low fields under applied pressure. Indeed, if we consider the magnetically ordered phase to be hosted within the field-polarized state, we can think of a continuity between the high *b*-axis field at ambient pressure and the zero-field state at high pressure. Whatever interactions are stabilized by field in the FP phase are also stabilized by pressure.

#### 4.3.2 Field approximately 25° from the *b* axis within the *bc* plane

The pressure-field-temperature phase diagram at this field angle has been constructed by complementary measurements of resistance and TDO frequency [100]. The field-pressure phase diagram for the lowest measured temperatures is shown in Figure 11(b).

At this field angle and at ambient pressure, the metamagnetic transition is coupled to the transition into $SC_{FP}$. As pressure is applied, the field of this coupled transition is suppressed monotonically. This is similar to the suppression of the field-polarized transition observed when H||*b*. However, unlike H||*b*, the upper critical field of SC1 is enhanced with pressure for field along this direction. The most striking effect of pressure application at this offset orientation is that at approximately 1 GPa, the borders of the SC1 and $SC_{FP}$ phases meet and resistivity remains zero from H = 0 T to at least 45 T at the lowest measured temperatures, as shown in Figure 11(b).

Moving to higher pressure, the transition to $SC_{FP}$ and the metamagnetic transition are clearly decoupled as a function of field. At the most extreme split, $H_m$ and the $SC_{FP}$ transition are separated by more than 20 T. At ambient and low pressure, $SC_{FP}$ initially seems reminiscent of URhGe or other reentrant actinide phases wherein ferromagnetic fluctuations parallel to H are essential to stabilizing high field superconductivity; however, the separation between these transitions suggests that magnetic fluctuations might not be responsible for pairing in high field $UTe_2$ [100].

The low-field superconducting state is always limited by $H_m$ in this case and does not cross over at any point with $SC_{FP}$. It is likely that the superconducting phases emerging from the paramagnetic state and $SC_{FP}$ are governed by different superconducting pairing mechanisms, unique to each phase [100]. Note that at approximately 1 GPa there is a transition between SC1 and SCP as distinguished by TDO measurements, but for simplicity this is not shown in Figure 11(b).

At 1.54 GPa, the field-polarized phase exhibits additional features in fields between the magnetically ordered and $SC_{FP}$ phases: as a function of field, there is both a broad peak in TDO frequency and a slight dip in magnetoresistance [100]. These features may represent a thermodynamically distinct phase, dubbed $A_{FP}$. The $A_{FP}$ anomalies only appear below 1.2 K, similar to the $SC_{FP}$ phase at this pressure, and below 1.2 K the temperature dependence of the two phases are similar [100]. This implies that the two phenomena are related, or at least that they occur on similar energy scales. Moreover, this indicates that $A_{FP}$ is distinct from the high-pressure magnetically ordered phase, which has a higher ordering temperature.

In consideration of the theoretical studies noted in Section 3.5.6, Ran *et al.* speculate that $A_{FP}$ and $SC_{FP}$ may be the same phase at different Landau levels. In this hypothesis, $SC_{FP}$ is enhanced from the partially superconducting (but never zero-resistance) $A_{FP}$ by the higher magnetic field, which increases the separation between Landau Levels and lessens the effects of energy-level broadening [100]. If this hypothesis is correct, then the resistance of $UTe_2$ should be equal to zero at approximately 90 T as the next Landau level would be reached.

### 4.3.3 Field along *a*

As can be seen in Figure 11(c), $H_{c2}$ for H || *a* gradually increases to its maximum of ~10 T just below 1 GPa, then gradually decreases to just above the ambient pressure value of ~6 T as $P_c$ is approached. The upper critical field for H || *a*, the magnetic easy axis, is still the lowest of the reported field orientations for all pressures below $P_c$ [20].

Figure 11(c) does not fully represent the complexity of the $UTe_2$ phase diagram for H || *a*, as the phase denoted “SC” in fact represents a number of different superconducting phases. The main plot of Figure 12 shows the temperature-dependence of $H_{c2}$ at select pressures below $P_c$, as determined from magnetoresistance [19]. As pressure increases, the effect of field on superconducting order becomes more pronounced. There is a kink in $H_{c2}$(T) at 0.6 K and 0.75 K for pressures ≈ 0.5 and 0.85 GPa, respectively, likely related to the presence of multiple

superconducting phases governed by different order parameters. AC calorimetry experiments have indicated the presence of up to four separate superconducting phases as a function of pressure and *a*-axis field [96]. As a further point of interest, the superconductivity of $UTe_2$ has reentrant behavior as a function of *a*-axis field at pressures near $P_c$. This can be clearly seen in the 1.4 GPa data in the main plot of Figure 12**Error! Reference source not found.** for temperatures near 2 K [19].

### 4.3.4 Field along *c*

For H || *c*, $H_{c2}(0)$ increases dramatically with the application of pressure, persisting to at least 27 T above 1 GPa as shown in Figure 11(d) [20]. Close to $P_c$, there is an apparently extreme reinforcement of superconductivity along the *c* axis, as indicated not only by this high $H_{c2}(0)$ but also by the almost vertical slope of $H_{c2}(T)$ at pressures near $P_c$, one example of which can be seen in the inset of Figure 12 **Error! Reference source not found.**[19, 20].

Near $P_c$ there is a region in which, at low temperatures, magnetic order seems to exist at low fields whereas superconductivity is stabilized at higher fields [20]. In this region there is hysteresis in the superconducting transition as a function of field, indicating a first-order transition between the superconducting and magnetically ordered phases [20]. Given the fact that the superconductivity is stabilized at high fields, Aoki *et al.* suggest that this is a phase similar to $SC_{FP}$ that emerges from a spin-polarized state. This may be related to reentrant superconductivity seen at lower fields in an off-axis field direction by Ran *et al.* [17]. By 1.61 GPa, superconductivity is entirely suppressed to at least 27 T [20]. Above $P_c$, a kink in resistivity has been observed which may be due to a WM state at fields above the M state [18].

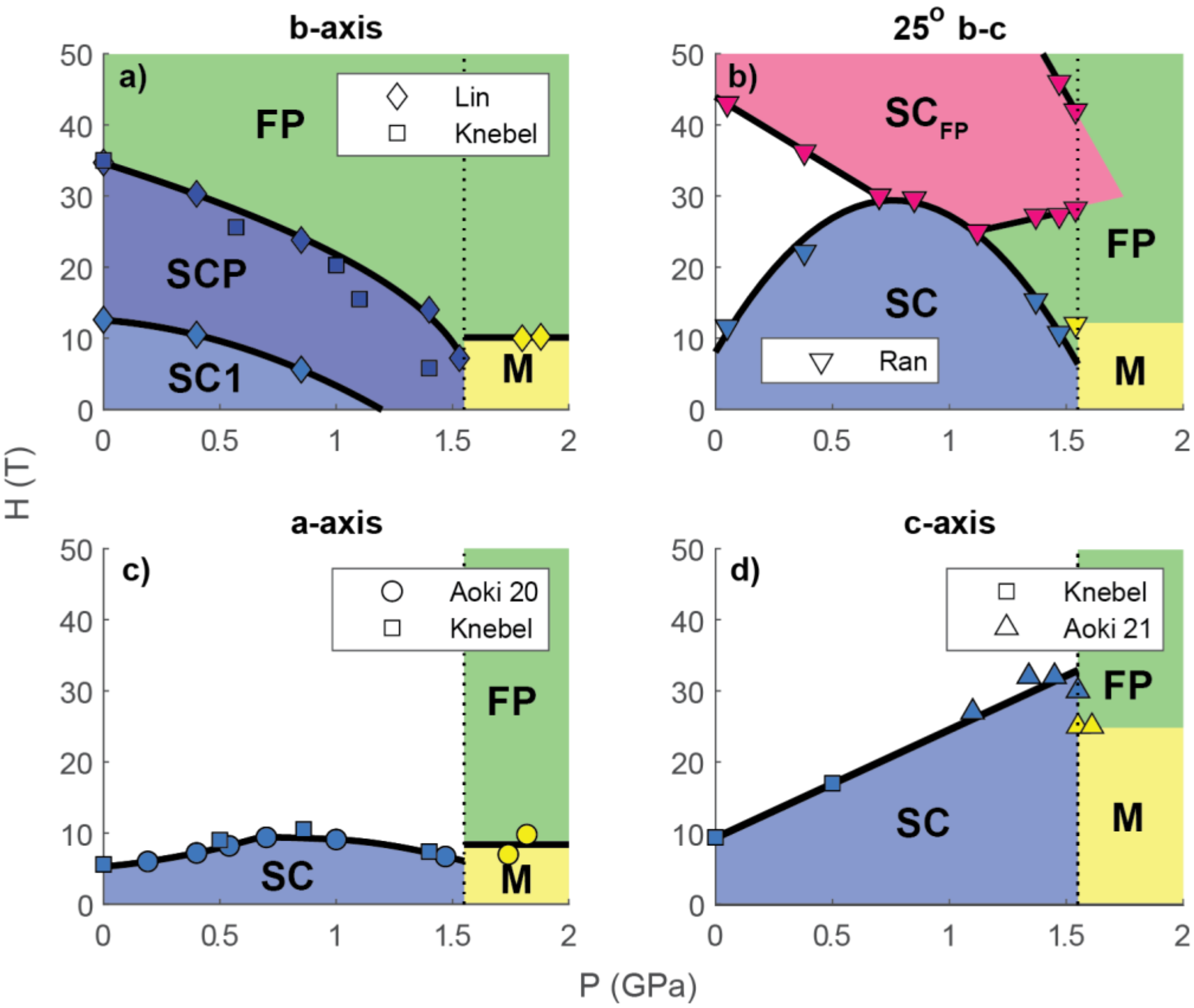

Figure 11. Pressure-field phase diagrams of $UTe_2$ with magnetic field directed (a) along the *b* axis, (b) within the *bc* plane, approximately 25° from the *b* axis, (c) along the *a* axis, and (d) along the *c* axis. Data are from Lin [16] (diamonds), Knebel [19] (squares), Ran [17] (downward triangles), Aoki [96] (circles), and Aoki [20] (upward triangles). Data displayed are from the lowest temperatures measured in each experiment. The labeled phases are described in the text.

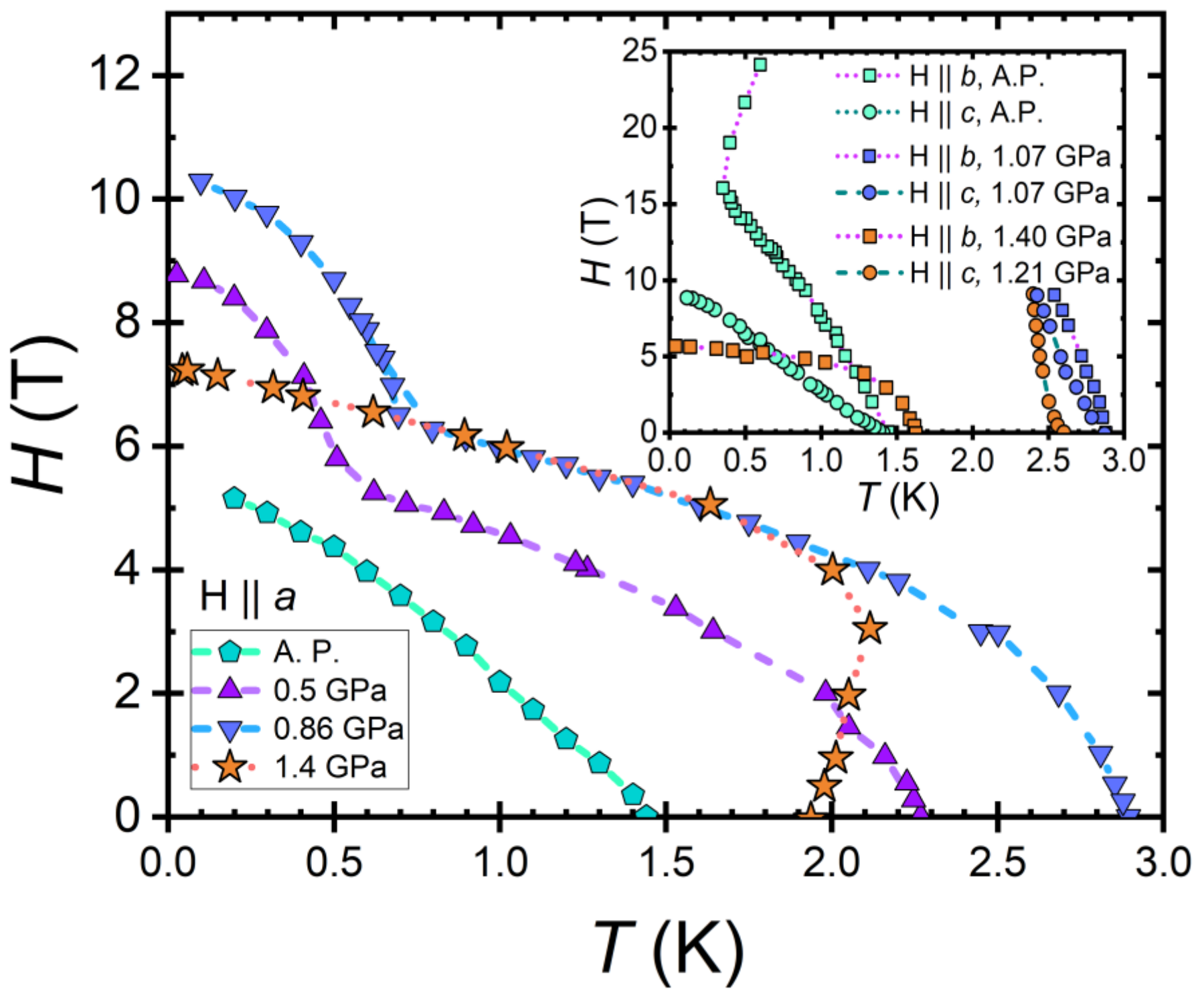

Figure 12. Evolution of $H_{c2}$ versus temperature with H || *a* for $UTe_2$ at ambient pressure (A.P.) (pentagons), 0.5 GPa (upward triangles), 0.86 GPa (downward triangles), and 1.4 GPa (stars). Inset shows H || *b* (squares) or H || *c* (circles) at A.P., close to 1 GPa, and near $P_c$. Data originally published by Knebel *et al.* [19].

## 5. Summary

Although research into this exciting material has only recently started, some points are becoming clearer with time. The main conclusions that we can draw so far are: 1) the zero-field superconducting state is spin-triplet, barring some unique mitigating factors to paramagnetic limitation such as a very small *g*-factor; 2) the enhanced *b*-axis superconductivity appears to be distinct from the lower-field superconductivity; 3) the high-field polarized phase is electronically distinct from the low-field normal state, possibly due to different Kondo hybridization, and hosts different superconducting states; and 4) the high-pressure electronic state is continuous with the state that is reached above the metamagnetic transition.

Many outstanding questions remain. First among them, the mechanisms that stabilize these different phases of superconductivity are a theoretical challenge. Measurements of the Fermi surface above the metamagnetic transition may yield insights into the nature of the $SC_{FP}$ phase. Experimentally, the possibility of superconductivity at even higher fields than the $SC_{FP}$ phase should be explored. Furthermore, the possible connection between the pressure-stabilized superconducting phase and the field-enhanced superconductivity with magnetic field along the *b* axis should be investigated, as should the nature of the pressure-induced magnetic order at zero field.

We therefore anticipate that $UTe_2$ will continue to be a fruitful topic of research for years to come, and that it will undoubtedly provide us with additional surprises.

## 6. Recent developments

While our manuscript has been under review, a great deal of work on $UTe_2$ has continued to be carried out and published.

A multi-component incommensurate charge density wave (CDW) has been observed in scanning tunneling microscopy, with an onset at low temperatures somewhere between 4 K and 10 K. The CDW intensity is suppressed with magnetic field along or near the (0 1 1) direction, and it seems to disappear approximately at the upper critical field of SC1 [101]. This apparent link between the CDW and superconductivity could arise naturally from a pair density wave state. Indeed, further scanning tunneling microscopy has uncovered the existence of a pair density wave in $UTe_2$ that shares the same wavevectors as the observed CDW [102].

Measurements of DC resistivity as a function of temperature and *b*-axis magnetic field have shown a region of low critical current in the phase diagram of $UTe_2$, attributed to a weakening in vortex pinning [103]. The lower-field boundary of this region coincides with subtle kinks in resistivity and AC magnetic susceptibility that were previously reported [62]; the authors of both manuscripts argue that these features demonstrates the existence of an intermediate superconducting state between SC1 and SC2.

Recently published x-ray scattering and spectroscopy results indicate that the uranium atoms in $UTe_2$ have an effective $5f^2 6d^1$ valence state, rather than the possible alternative $5f^3$ state [104].

The normal-state resistivity of $UTe_2$ has been studied as a function of field and temperature, for current along the *a* and *c* axes and for fields up to 70 T along all three principal crystallographic axes [105]. There is a clear anisotropy in the temperature dependence of the *a*-axis versus *c*-axis longitudinal resistivity, which the authors attribute to anisotropic scattering due to fluctuating magnetic moments. The measurements also indicate characteristic energy scales corresponding to approximately 15 K and 35 K in $UTe_2$, though the microscopic origin of these energy scales is unknown.

Studies of hydrostatic pressure on $UTe_2$ have now been performed to higher pressures than those reported in this review. At approximately 4 GPa, $UTe_2$ undergoes a structural phase transition from body-centered orthorhombic to body-centered tetragonal [106]. In this body-centered tetragonal phase, measurements of resistivity indicate that a new superconducting phase emerges around 7 GPa, with a maximum observed critical temperature of roughly 2 K [106]. Unlike the ambient pressure phases of $UTe_2$, this superconducting state has a relatively low upper critical field of only 2.5 T, with field applied along the *c* axis of the original orthorhombic unit cell.

In Section 2.4, we referred to de Haas-van Alphen (dHvA) oscillation measurements performed by Aoki *et al.* [39]. The observation of quantum oscillations was seemingly enabled by a new crystal growth technique involving a molten salt flux (MSF), the details of which have now been published [107]. Since this first observation, several other quantum oscillation studies have been performed on $UTe_2$, including further dHvA measurements using the field-modulation technique [108], dHvA measurements from magnetic torque [109], and measurements of oscillations in TDO frequency [110]. The experiments to date all confirm the existence of cylindrical Fermi surfaces along $k_z$. The TDO measurements showed frequencies that may be due to magnetic breakdown between cylindrical electron-like and hole-like Fermi surfaces. The TDO measurements also indicate the existence of a small, three-dimensional pocket of the Fermi surface, though the dHvA measurements show no indication of such a pocket; this discrepancy must be further investigated in the future.

Measurements of magnetic susceptibility have shown that MSF-grown crystals tend to have minor inclusions of ferromagnetic impurity phases such as $U_3Te_5$ [107, 111].

Crystals grown using the MSF method, compared to those grown by CVT, appear to have greater upper critical fields of the SC1 phase and an SC2 phase that persists to moderately higher angles of magnetic field away from the *b* axis [112]. However, the overall phase diagram remains qualitatively the same.

In terms of the FP and $SC_{FP}$ phases, the MSF-grown samples appear to have the same phase boundaries as previously measured CVT-grown samples [112]. It has also been reported that a CVT-grown sample with no SC1 or SC2 phase above 110 mK still exhibits an $SC_{FP}$ phase, though in a diminished region of phase space [113]. This counterintuitive result indicates that the $SC_{FP}$ phase is much more robust to disorder than the SC1 or SC2 phases, though the reason for this is not yet known.

The lower critical fields of the SC1 phase have been studied for MSF-grown crystals [114]. For fields along the *b* and *c* axes, the lower critical fields' temperature dependence deviates from predictions of Ginzburg-Landau theory, which the authors attribute to the influence of anisotropic ferromagnetic fluctuations.

The order parameter(s) of the SC1 phase remains under debate.

Measurements on a MSF-grown crystal showed a significant change in Knight shift when going through $T_c$ of the SC1 phase for fields along all three crystalline axes [115]. In contrast, previously reported NMR measurements of the SC1 state showed no change in Knight shift at $T_c$ for fields along the *a* axis [7]. The new findings have been interpreted to be consistent with a spin-triplet $A_u$ state rather than the authors' previously hypothesized spin-triplet $B_{3u}$ state.

TDO-based magnetic penetration depth measurements of both CVT-grown and MSF-grown $UTe_2$ samples indicate a multi-component order parameter for SC1, based on the temperature-dependence of the penetration depth [116]. From the anisotropy of the penetration depth, the authors suggest the gap state of SC1 has point nodes near the $k_y$- and $k_z$-axes, which they show could arise from a $B_{3u} + iA_u$ order parameter.

Scanning SQUID susceptometry of $UTe_2$ has been used to measure the temperature dependence of the superfluid density in the SC1 phase [117]. The results are consistent with either a $B_{3u}$ order parameter for a cylindrical Fermi surface or a $B_{1u}$ order parameter for a three-dimensional Fermi surface, with a highly anisotropic $A_u$ component allowed in either case. However, the authors note that the $A_u$ component must be much smaller than the second component to be consistent with their observations, in contrast with the conclusions of Ref. [116]. The scanning SQUID measurements found no evidence for a second phase transition for the SC1 phase.

New measurements of both CVT-grown and MSF-grown samples have shown a field-trainable polar Kerr effect, similar to the effect Ref. [10] previously reported [111]. The authors argue that this is attributable to local sample inhomogeneities rather than an intrinsic effect.

If the interpretation of Ref. [10] is correct, then a $B_{1g}$-like strain should be able to further split two nearly degenerate superconducting transitions. By applying such a strain and finding no noticeable split in the superconducting transition observed in heat capacity, Girod *et al.* showed that either the choices of order parameter from Ref. [10] are incorrect or the coupling between strain and the superconducting order parameters must be small [118].

Readers of this manuscript may also be interested in a recently published review on multiple superconducting phases in heavy-fermion metals, which includes detailed discussion of $UTe_2$ [119].